\documentclass[a4paper,11pt]{article}
\pdfoutput=1 
\usepackage{jheppub} 
\usepackage{tikz}
\usepackage{slashed}
\usepackage{ulem}
\usepackage{subfigure}
\usepackage{comment}
\usepackage{yhmath}
\usepackage[all]{xy}
\usepackage{multirow}
\usepackage{float}
\usepackage{mathtools}
\usepackage{xfrac}
\usepackage{wrapfig}
\usepackage{dsfont}
\usepackage{makecell}

\usepackage{color}

\newcommand \be{\begin{eqnarray}}
\newcommand \ee{\end{eqnarray}}

\numberwithin{equation}{section}

\DeclareMathOperator{\Tr}{Tr}

\DeclareMathOperator{\diag}{diag}

\newcommand{\bea}{\begin{eqnarray}}
\newcommand{\eea}{\end{eqnarray}}
\newcommand{\beq}{\begin{equation}}
\newcommand{\eeq}{\end{equation}}
\newcommand{\bal}{\begin{equation}\begin{aligned}}
\newcommand{\eal}{\end{aligned} \end{equation}}

\newcommand{\vev}[1]{{\left< {#1} \right>}}

\newcommand{\cA}{{\mathcal A}}

\newcommand{\cL}{{\mathcal L}}

\newcommand{\cN}{{\mathcal N}}
\newcommand{\cP}{{\mathcal P}}
\newcommand{\cQ}{{\mathcal Q}}
\newcommand{\cT}{{\mathcal T}}

\newcommand{\cO}{{\mathcal O}}

\newmuskip\pFqmuskip

\newcommand*\pFq[6][8]{%
  \begingroup 
  \pFqmuskip=#1mu\relax
  \mathchardef\normalcomma=\mathcode`,
  \mathcode`\,=\string"8000
  \begingroup\lccode`\~=`\,
  \lowercase{\endgroup\let~}\pFqcomma
  {}_{#2}F_{#3}{\left[\genfrac..{0pt}{}{#4}{#5};#6\right]}%
  \endgroup
}
\newcommand{\pFqcomma}{{\normalcomma}\mskip\pFqmuskip}

\title{Framing fermionic Wilson loops in ABJ(M)}

\author[a]{Marco S. Bianchi,}
\author[b]{Luigi Castiglioni,}
\author[b]{Silvia Penati,}
\author[b]{Marcia Tenser,}
\author[c,d]{Diego Trancanelli}

\affiliation[a]{Facultad de Ingeniería, Arquitectura y Diseño, Universidad San Sebastián,
Santiago, Chile}
\affiliation[b]{Dipartimento di Fisica, Universit\`a degli Studi di Milano--Bicocca and INFN, Sezione di Milano--Bicocca, Piazza della Scienza 3, 20126 Milano, Italy}
\affiliation[c]{Dipartimento di Scienze Fisiche, Informatiche e Matematiche, Universit\`a di Modena e Reggio Emilia, via G. Campi 213/A, 41125 Modena, Italy}
\affiliation[d]{INFN Sezione di Bologna, via Irnerio 46, 40126 Bologna, Italy}

\emailAdd{marco.bianchi@uss.cl}
\emailAdd{l.castiglioni8@campus.unimib.it}
\emailAdd{silvia.penati@mib.infn.it}
\emailAdd{marciatenser@gmail.com}
\emailAdd{diego.trancanelli@unimore.it} 

\abstract{
\noindent
Framing plays a central role in the evaluation of Wilson loops in theories with Chern-Simons actions. In pure Chern-Simons theory, it guarantees topological invariance, while in theories with matter like ABJ(M), our theory of interest, it is essential to enforce the cohomological equivalence of different BPS Wilson loops. This is the case for the 1/6 BPS bosonic and the 1/2 BPS fermionic Wilson loops, which have the same expectation value when computed as matrix model averages from localization. This equivalence holds at framing $\mathfrak{f}=1$, which has so far been a challenge to implement in perturbative evaluations. In this paper, we compute the expectation value of the 1/2 BPS fermionic circle of ABJ(M) theory up to two loops in perturbation theory {\it at generic framing}. This is achieved by a careful analysis of fermionic Feynman diagrams, isolating their framing dependent contributions and evaluating them in point-splitting regularization using framed contours. Specializing our result to $\mathfrak{f}=1$ we recover exactly the matrix model prediction, thus realizing for the first time a direct perturbative check of localization for this operator. We also generalize our computation to the case of a multiply wound circle, again matching the corresponding matrix model prediction.
}

\keywords{Chern-Simons theories, Wilson, 't Hooft and Polyakov loops, Topological Field Theories}
\arxivnumber{}

\begin{document}
\maketitle


\let\clearpage\relax
\allowdisplaybreaks
\section{Introduction}

Since the seminal paper by Witten on knot theory \cite{Witten:1988hf}, it has been clear that framing plays a crucial role in the definition of topological invariants of three-manifolds via Chern-Simons theory. In fact, the functional quantization of the theory unavoidably requires the introduction of a metric, thus breaking general covariance. Similarly, a na\"ive regularization of gauge invariant operators like Wilson loops does not lead to topologically invariant quantities in the quantum theory. Both in the partition function and in the Wilson loops, a topological anomaly arises as a non-invariant phase factor. As discussed in \cite{Witten:1988hf}, one can then restore topological invariance by compensating with a {\em framing phase} counterterm. In the partition function it originates by choosing a particular trivialization of the tangent bundle of the three-manifold where the theory is defined. In the evaluation of Wilson loops it can be traced back to a particular choice of point-splitting regularization of short distance singularities \cite{Witten:1988hf,Guadagnini:1989am,Alvarez:1991sx}. Topological invariance is thus recovered at the price of introducing a non-trivial scheme dependence. However, this is a controlled dependence, as it is clear how topologically invariant quantities change under a change of scheme. 

Notoriously, pure Chern-Simons theory may be supersymmetrized by the introduction of some quiver structure with a Chern-Simons gauge field at each node and matter fields in the (anti)bifundamental representation connecting the nodes. This results in $2\leq {\cal N}\leq 6$ Chern-Simons-matter theories, with ABJ(M) theory \cite{Aharony:2008ug,Aharony:2008gk} being the maximally supersymmetric one. It has a 2-node quiver with a $U(N_1)$ gauge field $A_\mu$ at the first node and a $U(N_2)$ gauge field $\hat A_\mu$ at the second node, with opposite Chern-Simons levels $\pm k$, $k >2$. It is coupled to four complex scalars $C_I$ and four fermions $\psi_I^\alpha$ ($I=1,2,3,4$ is an $SU(4)_R$ R-symmetry index). This theory is holographically dual to M-theory on $AdS_4 \times S^7/\mathbb{Z}_k$, or, for large $k$, to type IIA superstrings on $AdS_4\times \mathbb{CP}^3$.

Because of the coupling to matter, ABJ(M) theory is no longer topological. Still, it has a topological sector of operators whose correlation functions do not depend on the insertion points \cite{Binder:2019mpb,Gorini:2020new,Armanini:2024kww}. More importantly for us, framing is still relevant in this theory when computing expectation values of non-local operators such as Wilson loops.

ABJ(M) theory has a wealth of BPS Wilson loops preserving different amounts of supersymmetry, see \cite{Drukker:2019bev} for a rather recent review. Salient representatives are the 1/6~BPS bosonic circle introduced in \cite{Drukker:2008zx,Chen:2008bp,Rey:2008bh} and the 1/2 BPS fermionic circle of \cite{Drukker:2009hy}. As the names suggest, the former operator is invariant under 4 supercharges, and is defined solely in terms of the bosonic fields of the theory,  {\it i.e.} the gauge fields and the complex scalars $C_I$. The latter operator also includes a coupling to the fermions $\psi^\alpha_I$ and preserves 12 supercharges. Given the premise above, it is quite natural to wonder whether framing plays any role in the computation of these operators. 

It has been known for quite some time that the answer to this question is in the affirmative. The reason is that BPS Wilson loops in ABJ(M) are amenable to evaluation via supersymmetric localization \cite{Kapustin:2009kz,Marino:2009jd} and that this procedure has a preferred framing choice \cite{Drukker:2010nc}, namely $\mathfrak{f}=1$. The ABJ(M) partition function localizes in fact to a finite-dimensional matrix model integral for two interacting sets of eigenvalues corresponding to the two nodes of the quiver and the expectation value of BPS Wilson loops is obtained by considering certain insertions in this integral. 

The 1/6 BPS bosonic and the 1/2 BPS fermionic Wilson loops are known to be cohomologically equivalent with respect to the supercharge ${\cal Q}$ used for localization \cite{Drukker:2009hy}. This means that they differ by a ${\cal Q}$-exact term and, despite being different classical operators, they correspond to the same matrix model insertion, as long as ${\cal Q}$ is not broken by quantum effects. It is known that this condition is satisfied at framing $\mathfrak{f}=1$ \cite{Kapustin:2009kz}, where indeed the two operators computed as a matrix model average have the same expectation value. 

However, when computed in ordinary perturbation theory, which is typically performed at zero framing, the expectation values of these cohomologically equivalent operators are found to be different \cite{Bianchi:2014laa,Bianchi:2016yzj,Bianchi_2016}. As we discuss in this paper, this is due to the appearance of a {\em cohomological anomaly} which breaks the equivalence between the two operators at framing zero. Since this anomaly turns out to be an overall phase, it can be compensated by a framing counterterm, following a prescription similar to the one used for restoring general covariance in Chern-Simons topological invariants \cite{Witten:1988hf}. According to the matrix model result, the cohomological anomaly is exactly canceled by the framing phase at $\mathfrak{f}=1$.\footnote{The same discussion also applies to other cohomologically equivalent operators, as for example the family of interpolating Wilson loops considered in \cite{Castiglioni:2022yes}. Their expectation value in perturbation theory depends on the interpolating parameters, even though these do not appear in the matrix model insertion. The mismatch, also in that case, can be fixed by framing phases relating the two different framing prescriptions.} 

In Chern-Simons theories with matter, the appearance of the framing phase is not just an {\it ad hoc} trick to force agreement between different procedures, rather it captures interesting physics. For example, it enters the prescription for computing the Bremsstrahlung function of the 1/2 BPS Wilson loop in ABJM theory exactly \cite{Bianchi:2014laa}. Therefore, a deeper interpretation of the framing phase is mandatory. Moreover, recalling that 1/6 BPS and 1/2 BPS Wilson loops define one-dimensional superconformal defect theories, it would be interesting to have a physical interpretation of framing directly from the point of view of the defect theory. Still, it would be desirable to be able to perform perturbative computations at generic framing to determine the phase interpolating between the perturbative results and the localization prediction. This would provide a strong test of the matrix model result and a solid basis for attempting a holographic interpretation of framing. 

A perturbative computation at generic framing was carried out for the $1/6$ BPS bosonic loop in \cite{Bianchi_2016}, but so far it has eluded efforts  in the case of fermionic operators. This is mainly due to technical difficulties in the evaluation of integrals coming from Feynman diagrams with fermions. In this paper, we manage to overcome these problems and compute the expectation value of the 1/2 BPS fermionic circle in ABJ(M) at generic framing, up to two-loop order.\footnote{In a different setup, the perturbative computation of mesonic Wilson lines ending on fundamental matter, both bosonic and fermionic, was carried out in \cite{Gabai:2022vri,Gabai:2022mya}. Also in that case framing played a central role in the analysis.} 

Our general strategy relies on separating the Feynman diagrams into contributions which are framing-independent and pieces potentially sourcing framing effects. The former were already analyzed in \cite{Bianchi:2013zda,Bianchi:2013rma,Griguolo:2013sma}, where the calculation was approached at vanishing framing up to two loops. Likewise, bosonic diagrams are fully under control, since their framing-dependent and independent contributions are known from the literature on pure Chern-Simons theory \cite{Guadagnini:1989am}. The novelty of our analysis lies in explicitly showcasing that fermionic diagrams of the 1/2 BPS Wilson loop in ABJ(M) theory are sensitive to framing and in calculating their contribution. This is achieved by considering slightly deformed contours from the circle and allowing field insertions on them, as in a point-splitting regularization of coincident operators. We then identify terms in the fermionic diagrams' integrands that are proportional to the Gauss linking integral, or slight modifications of it.  These are sensitive to the linking number between the deformed contours, which survives the removal of the small deformations, though not in a topologically invariant manner, contrary to pure Chern-Simons theory.

Our calculation provides a firmer test of localization than previous works \cite{Bianchi:2013rma,Griguolo:2013sma}, as it includes framing effects at all steps along the way. Moreover, our analysis is performed for arbitrary framing and is further generalized to Wilson loops with multiple windings. These results clarify the role of framing in perturbative calculations of Wilson loops and its relation to localization. They also lay the bases for a deeper understanding of how cohomological equivalences within families of supersymmetric Wilson loops are realized perturbatively.

The paper is organized as follows. In section \ref{sec:theory} we review some basic notions about the Wilson loops of ABJ(M) theory and their evaluation as matrix model averages via localization. In section \ref{sec:framing} we discuss framing and its role in guaranteeing the cohomological equivalence of different BPS operators. Section \ref{sec:circWLperturbative} is the core of the paper, in which we evaluate, for the first time, fermionic Feynman diagrams at generic framing. This allows us to verify the matrix model prediction for the expectation value of the 1/2 BPS fermionic Wilson loop, up to second order in perturbation theory, directly at framing $\mathfrak{f}=1$. Finally, in section \ref{sec:multiplewindings} we repeat the computation for multiply wound loops, providing a non-trivial check of our method. Technical details are collected in a series of appendices.
\section{Preliminaries}
\label{sec:theory}

We start by reviewing some basic notions about supersymmetric Wilson loops in ABJ(M) theory, such as their definition and the computation of their expectation value using localization to a matrix model.


\subsection{BPS Wilson loops in ABJ(M) theory}
\label{sec:WLsABJM}

In $\cN=6$ $U(N_1)_k \times U(N_2)_{-k}$ ABJ(M) theory \cite{Aharony:2008ug,Aharony:2008gk}, one can define BPS Wilson loops preserving different amounts of supersymmetry, see \cite{Drukker:2019bev} for a review. This is achieved by including couplings to the matter fields in the holonomies of $A_{\mu}$ and $\hat A_{\mu}$ at the two nodes of the quiver. 

A first possibility is to add bilinears of the scalar fields $C_I$ and $\bar C^I$ through a scalar coupling matrix $M_J^{\ I}$, where $I,J=1,2,3,4$ are $SU(4)$ R-symmetry indices \cite{Drukker:2008zx,Chen:2008bp}. This is possible because in three dimensions the scalars have dimension $1/2$ and a bilinear of (anti)bifundamental fields transforms in the adjoint representation of the gauge groups of the nodes, as $A_{\mu}$ and $\hat A_{\mu}$ do. This gives rise to so-called bosonic Wilson loops of the form\footnote{Throughout the paper we only consider Wilson loops in the fundamental representation.}
\begin{equation}\label{eq:Wbos}
\begin{split}
    W^\textrm{bos}  &= 
    \Tr \cP \exp\left(-i \oint \cA \,d\tau \right)\,,\qquad \cA = A_\mu \dot{x}^\mu - \frac{2 \pi i}{k}\vert\dot{x}\vert M_J^{\ I} C_I \bar{C}^J\,, \\
    \hat W^\textrm{bos} &= 
    \Tr \cP \exp\left(-i \oint \hat\cA \,d\tau \right)\,,\qquad \hat\cA = \hat{A}_\mu \dot{x}^\mu - \frac{2 \pi i}{k}\vert\dot{x}\vert M_J^{\ I} \bar{C}^J C_I\,.
\end{split}
\end{equation}
For a contour along a unit radius circle in $\mathbb{R}^3$,
\begin{equation}\label{eq:maximalcircle}
    x^\mu=(\cos\tau,\sin\tau,0)\,, \quad 0 \leq \tau \leq 2\pi\,,
\end{equation}
the choice $M_I^J = \text{diag}(-1,-1,1,1)$ preserves 4 supercharges, giving 1/6 BPS bosonic Wilson loops \cite{Drukker:2008zx,Chen:2008bp}. Taking into account the quiver structure of the theory, \eqref{eq:Wbos} can be recast in terms of a single operator that couples to both nodes and still preserves 1/6 of the supercharges
\bea
\label{eq:bosonicWL1}
    W_{1/6} = \Tr \cP \exp\left(-i\oint \cL_0 \,d\tau\right) 
    \,,\qquad \cL_0 = \begin{pmatrix}
        \cA & 0 \\ 0 & \hat\cA
    \end{pmatrix},
\eea
where ${\cal L}_0$ is a $U(N_1) \times U(N_2)$  connection.

More generally, $\cL_0$ can be promoted to a $U(N_1|N_2)$ superconnection ${\cal L}$ by adding fermions in its off-diagonal entries \cite{Drukker:2009hy}.\footnote{See also \cite{Berenstein:2008dc,Lee:2010hk} for the construction of this operator from a Higgsing procedure.} This defines what is called a fermionic Wilson loop 
\begin{equation}\label{eq:fermioniWL1}
   W  = \frac{\text{STr}\left( \cP e^{-i\oint \cL d\tau }\cT \right)}{\text{STr}(\cT) }\,,\qquad \cL = \begin{pmatrix}
        \cA & -i\sqrt{\frac{2\pi}{k}}|\dot x| \eta_I \bar \psi^I \\ -i\sqrt{\frac{2\pi}{k}}|\dot x|  \psi_I \bar\eta^I & \hat\cA
    \end{pmatrix}\,.
\end{equation}
The couplings $\eta_I$ and $\bar\eta^I$ are commuting spinors and the twist matrix $\cT$ ensures (super)gauge invariance \cite{Drukker:2009hy, Cardinali:2012ru}.\footnote{It is possible to define $W$ without the insertion of the twist matrix ${\cal T}$, at the price of introducing constant shifts in the diagonal blocks, see chapter 2 of \cite{Drukker:2019bev}. While the latter is a more elegant formulation, it complicates the perturbative analysis, see chapter 5 of \cite{Drukker:2019bev}. For this reason, in this paper we have opted to keep the twist matrix.} Depending on the contour, a suitable choice of $\cT$, $M_J^{\ I}$ and $\eta_I,\bar\eta^I$ may render the resulting operator supersymmetric. Notice that, unlike what happens with the ${\cal L}_0$ above, the supersymmetry variation of ${\cal L}$ is not zero, it is rather given by a total covariant derivative
\bea
\delta {\cal L}= {\cal D}_\tau {\cal G} \equiv \partial_\tau {\cal G}+i\{{\cal L},{\cal G}],
\eea
where ${\cal G}$ is some $U(N_1|N_2)$ supermatrix. In this way, the (super)gauge invariance of the operator guarantees also the invariance under supersymmetry transformations.

We are interested in the maximally supersymmetric operator, the $1/2$ BPS fermionic Wilson loop of \cite{Drukker:2009hy}, which we call $W_{1/2}$.
This is obtained by taking the contour along the circle in \eqref{eq:maximalcircle}, the twist matrix to be $\cT=\diag(1_{N_1},-1_{N_2})$, the scalar coupling matrix as $M_I^J = \text{diag}(-1,1,1,1)$, and the fermionic couplings as
\begin{equation}\label{eq:roadmapcouplingswithphases}
    \eta_I^{\alpha} = 2e^{i\tau/2}\delta_I^1 \left(\bar{s}\, \Pi_+\right)^{\alpha} \,, \qquad \bar\eta^I_{\alpha} = 2ie^{-i\tau/2}\delta^I_1 \left(\Pi_+ \,s \right)_{\alpha}\,,
\end{equation}
where $\Pi_+$ is the projector
\begin{equation}\label{eq:projector}
    \Pi_+ \equiv \frac{1}{2} \left( 1 + \frac{\gamma\cdot \dot{x}}{|\dot{x}|} \right)\,,
\end{equation}
$\gamma^\mu$ are the Pauli matrices, and $s_\alpha,\bar{s}_\alpha$ are constant commuting spinors\footnote{Contracted spinor indices are omitted assuming $\xi\psi\equiv \xi^\alpha\psi_\alpha$. Spinor indices are raised as $\xi^\alpha=\epsilon^{\alpha\beta}\xi_\beta$, with $\epsilon^{+-}=1$.}
\begin{equation}\label{eq:sspinors}
    s_\alpha = \begin{pmatrix}
        1 \\ 0
    \end{pmatrix}_\alpha\,, \qquad \bar{s}_\alpha = \begin{pmatrix}
        0 \\ 1
    \end{pmatrix}_\alpha\,.
\end{equation}
As the name says, this operator preserves 12 supercharges.

Details about this specific parametrization of the matter couplings can be found in chapter 2 of \cite{Drukker:2019bev}. The important thing to note is that, apart from the overall phases $e^{\pm i \tau/2}$ appearing in \eqref{eq:roadmapcouplingswithphases},\footnote{These phases can be eliminated by a gauge transformation \cite{Drukker:2019bev} and do not play a relevant role for framing. For example, when they enter the Gauss linking integral, they do not change its value, as proven in appendix \ref{app:calugareanu}. } the rest of the couplings are defined in terms of the embedding coordinates of the contour $x^\mu(\tau)$, rather than the parameter $\tau$ itself. This is going to be important in section \ref{sec:circWLperturbative} in order to be able to implement framing and to consider contours that deviate from the circle.

Many different generalizations of the operators above are possible. For example, while maintaining a circular contour, one could deform the superconnection ${\cal L}$ by dressing it with appropriate parameters, which allow to interpolate between different Wilson loops \cite{Drukker:2019bev, Drukker:2020dvr}. These parameters act as marginally relevant deformations that trigger non-trivial RG flows connecting the operators described above \cite{Castiglioni:2022yes,Castiglioni:2023uus, Castiglioni:2023tci}. In this paper we do not consider interpolating operators (whose analysis is left for future studies) and we focus instead on the fixed points of such RG flows, in particular on the $1/6$ BPS bosonic and the $1/2$ BPS fermionic Wilson loops.  

Another possibility is to choose more general contours \cite{Cardinali:2012ru}, instead of working with the  circle \eqref{eq:maximalcircle}. One notable example is to the so-called latitude, which is supported on
\begin{equation}\label{eq:latitudecontour}
    x^{\mu} = ( \cos\theta_0\cos\tau,\cos\theta_0\sin\tau, \sin\theta_0)\,, \quad \theta_0 \in [-\tfrac{\pi}{2}, \tfrac{\pi}{2}]\,,
\end{equation}
with $\theta_0$ being the latitude angle. In this case the maximally supersymmetric Wilson loop is the 1/6 BPS fermionic operator \cite{Bianchi:2014laa}, which corresponds to choosing the couplings as
\begin{equation}
\label{eq:latitudecouplings}
    M_J^{\ I} = \begin{pmatrix} -\nu && 0 && e^{-i\tau}\sqrt{1-\nu^2} && 0 \\
    0 && 1 && 0 && 0 \\
    e^{i\tau}\sqrt{1-\nu^2} && 0 && \nu && 0 \\
    0 && 0 && 0 && 1
    \end{pmatrix}\,,\quad \eta_I^{\alpha} = \frac{e^{\frac{i\nu\tau}{2}}}{\sqrt{2}} \begin{pmatrix}\!\!\!\!
    \sqrt{1+\nu}\!\!\!\! \\ \!\!\!\! -\sqrt{1-\nu}e^{i\tau}\!\!\!\! \\ 0 \\ 0
    \end{pmatrix}_I \!\!\!\! \begin{pmatrix}
        1, -i e^{-i\tau}
    \end{pmatrix}^{\alpha}\,,
\end{equation}
with $\nu \equiv \cos\theta_0$, and the twist matrix
\begin{equation}
\label{eq:twistmatrixlatitude}
    \cT = \begin{pmatrix}
        e^{-\frac{i\pi\nu}{2}} & 0 \\ 0 & e^{\frac{i\pi\nu}{2}} 
    \end{pmatrix}\,.
\end{equation}

While we focus mostly on the circle, in section \ref{sec:circWLperturbative} we comment on how framing works for the latitude, evaluating a certain Feynman diagram that serves as a consistency check of our method. 


\subsection{The matrix model computation}
\label{sec:matrixmodel}

Using supersymmetric localization, the partition function of ABJ(M) theory can be reduced to the non-Gaussian matrix model \cite{Kapustin:2009kz}
\begin{equation}\label{eq:MM}
 Z = \int \prod_{a=1}^{N_1} d\lambda_a \,e^{i\pi k\lambda_a^2} \prod_{b=1}^{N_2} d\hat\lambda_b \,e^{i\pi k\hat\lambda_b^2} \,\frac{\prod_{a<b}^{N_1} \sinh^2{\left(\pi(\lambda_a-\lambda_b)\right)} \prod_{a<b}^{N_2} \sinh^2{\left(\pi(\hat\lambda_a-\hat\lambda_b)\right)}}{\prod_{a=1}^{N_1} \prod_{b=1}^{N_2} \cosh^2{\left(\pi(\lambda_a-\hat\lambda_b)\right)}}
\end{equation}
for two sets of eigenvalues $\lambda_a$ and $\hat\lambda_a$ associated with the two nodes of the quiver.

The evaluation of the expectation value of supersymmetric Wilson loops as a matrix model average amounts to inserting appropriate functions of the eigenvalues in \eqref{eq:MM} \cite{Marino:2009jd,Drukker:2009hy,Kapustin:2009kz}. For the circular loops $W^\text{bos}$ and $\hat{W}^\text{bos}$ in \eqref{eq:Wbos}, the insertions are
\begin{equation}\label{eq:MM16}
 w^\text{bos}= \frac{1}{N_1} \sum_{a=1}^{N_1} e^{2\pi\lambda_a}\,,\quad \hat{w}^\text{bos}= \frac{1}{N_2} \sum_{a=1}^{N_2} e^{2\pi\hat\lambda_a}\,,
\end{equation}
respectively, whereas for $W_{1/6}$ it is
\begin{equation}\label{eq:MM12}
  w_{1/6}= \frac{1}{N_1+N_2} \left( \sum_{a=1}^{N_1} e^{2\pi\lambda_a}+ \sum_{a=1}^{N_2} e^{2\pi\hat\lambda_a} \right)\,.
\end{equation}

Interestingly, the 1/2 BPS fermionic Wilson loop is also captured by the same matrix model insertion. This is due to the fact that this operator and the 1/6 BPS bosonic Wilson loop are cohomologically equivalent, differing solely by a BRST-exact term \cite{Drukker:2009hy}
\begin{equation}
\label{eq:cohomo}
W_{1/2} = W_{1/6} + {\mathcal Q}V,
\end{equation}
where ${\mathcal Q}$ is a linear combination of the supercharges mutually preserved by the two operators, and $V$ is a supermatrix that can be explicitly determined. This cohomological equivalence is not exclusive to this pair of Wilson loops. In fact, it has been proved to hold in more general classes of operators, from the 1/6 BPS fermionic operators \cite{Ouyang:2015bmy} to the latitude \cite{Bianchi:2014laa} and the more general 1/24 and 1/12 BPS parametric operators introduced in~\cite{Castiglioni:2022yes}. 

Consequently, the expectation values of cohomologically equivalent operators, as computed localizing the functional integral with ${\cal Q}$, must all coincide. More precisely, one expects that matrix model (mm) results satisfy
\begin{equation}\label{eq:W12f1}
 \langle W_{1/2} \rangle_\textrm{mm} = \langle W_{1/6} \rangle_\textrm{mm} = \frac{N_1 \langle W^\text{bos}\rangle_\textrm{mm} + N_2 \langle \hat{W}^\text{bos}\rangle_\textrm{mm}}{N_1+N_2}\,.
\end{equation}
Expanding the matrix model computation at small couplings one finds \cite{Marino:2009jd,Drukker:2010nc} 
\begin{equation}
\begin{split}
& \langle W^\text{bos} \rangle_\textrm{mm} =  1 + \frac{i\pi N_1}{k}- \frac{\pi^2}{6k^2}\left( 
    4N_1^2-6N_1N_2 -1 \right) + \mathcal{O}\left(k^{-3}\right)\, , \\
    & \langle \hat{W}^\text{bos} \rangle_\textrm{mm} =  1 - \frac{i\pi N_2}{k}- \frac{\pi^2}{6k^2}\left( 
    4N_2^2-6N_1N_2 -1 \right) + \mathcal{O}\left(k^{-3}\right)\, , 
    \end{split}
    \end{equation}
and therefore
\begin{equation}\label{eq:MMresult}
    \langle W_{1/2} \rangle_\textrm{mm}  = 1 + \frac{i\pi(N_1-N_2)}{k}- \frac{\pi^2}{6k^2}\left[ 
    4(N_1^2+N_2^2)-10N_1N_2-1 \right] + \mathcal{O}\left(k^{-3}\right)\,.
\end{equation}

A direct perturbative computation of $\vev{W_{1/2}}$ has been performed, up to second order in perturbation theory, in \cite{Bianchi:2013zda,Bianchi:2013rma,Griguolo:2013sma}. Remarkably, it does not immediately match the expression above, but it has to be corrected {\it a posteriori} by the introduction of a phase. The reason is that the matrix model selects a preferred framing \cite{Marino:2009jd,Drukker:2010nc}, specifically $\mathfrak{f}=1$, while perturbative analyses are generically performed at zero framing.

In the next section we discuss the role of framing in the computation of Wilson loops in theories with Chern-Simons terms and then, in section \ref{sec:circWLperturbative}, we set out to reproduce \eqref{eq:MMresult} directly from a perturbative computation done at generic framing and then specialized at~$\mathfrak{f}=1$.
\section{Framing in Wilson loops}
\label{sec:framing}


\subsection{Pure Chern-Simons theory}

Framing was first introduced in pure Chern-Simons (CS) theories in Witten's seminal paper on knot theory \cite{Witten:1988hf}. It was needed as a regularization required  to remove a topological anomaly which unavoidably arises in the partition function due to the gauge-fixing procedure. More precisely, a topologically invariant partition function can be defined by trivializing the tangent bundle of the three-dimensional manifold, including a phase counterterm to cancel the anomaly. The scheme dependence introduced by the counterterm ({\it i.e.} by the trivialization) is encoded into an integer $\mathfrak{f}$ which is called {\em framing}. The appearance of such a scheme dependence is not a problem, as the partition function changes in a controlled way under a change of scheme: at large $k$, it simply gains a phase $e^{i\pi \mathfrak{f}\frac{\textrm{dim}(G)}{12}}$, for a gauge group $G$. We refer to \cite{mooretasi} for a recent review. 

For analogous reasons, the expectation value of Wilson loops in Chern-Simons theories turns out to be a metric dependent quantity. However, topological invariance can be restored by framing the contour \cite{Witten:1988hf}. In quantum perturbation theory this can be seen as a point-splitting regularization of short distance singularities.  Framing is then the scheme-dependent phase which survives after removing the regulator. This interpretation as point-splitting explains why framing appears also in Chern-Simons theories with matter, like ABJ(M), which are not topological. 

For pure Chern-Simons theory the first evaluation of the expectation value of Wilson loops at two loops for generic framing was performed in \cite{Guadagnini:1989am}. Let us review the main idea behind the computation. Along a smooth path $\Gamma$, the expectation value reads
\begin{equation}
\label{eq:VEV}
    \langle W_\text{CS}\rangle = \frac{1}{N}\langle \Tr\cP e^{-i\int_{\Gamma} A_{\mu} dx^{\mu}} \rangle\,.
\end{equation}
The perturbative expansion of the operator produces path-ordered integrated $n$-point functions of the form 
\begin{equation}
\int_{\tau_1 > \tau_2 > \dots > \tau_n} \!\! \!\!\!\langle A_{\mu_1}(x_1) A_{\mu_2}(x_2) \ldots A_{\mu_n}(x_n) \rangle \, \dot{x}_1^{\mu_1} \dots \dot{x}_n^{\mu_n} \, d\tau_1 \dots d\tau_n \, , 
\end{equation}
where $x_i\equiv x(\tau_i)$. Potential singularities arising from contractions of fields at coincident points can be regularized via a point-splitting procedure, that is by moving the $x_i$'s to auxiliary contours, infinitesimally displaced from the original path $\Gamma$,
\begin{equation}\label{eq:framedcontour}
   x^{\mu}_i \to  x^{\mu}_i + \delta\, (i-1)\, n^{\mu}(\tau_i)\,,
   \qquad |n(\tau_i)|=1\,,
\end{equation}
where $\delta$ is a small deformation parameter and $n^{\mu}(\tau_i)$ are vector fields orthonormal to $\Gamma$. 

For instance, the one-loop correction to the expectation value \eqref{eq:VEV} is proportional to the integrated gauge propagator $\vev{A_{\mu_1}(x_1)A_{\mu_2}(x_2)}$ (see \eqref{eqn:propagator}), leading to
\begin{equation}\label{eq:CS1loop}
    \int_{\Gamma}dx_1^{\mu} \int_{\Gamma} dx_2^{\nu} \, \epsilon_{\mu\nu\rho} \frac{(x_1-x_2)^{\rho}}{|x_1-x_2|^3}\,.
\end{equation}
Although the propagator can be proved to be contractible to a point, implying that the integral does not develop a singularity at $x_1 = x_2$, we still apply the framing procedure to compute it. We let $x_1$ run on the original contour $\Gamma$, while $x_2$ runs on an infinitesimal deformation of the original path
\begin{equation}
\label{x2contour}
   \Gamma_\mathfrak{f} :  x_2^{\mu} \to x_2^{\mu}+\delta \, n^{\mu}(\tau_2) \,, \qquad |n(\tau_2)|=1\,.
\end{equation}
The resulting contribution is proportional to the Gauss linking integral 
\begin{equation}
\label{eq:linking}
    \chi(\Gamma,\Gamma_\mathfrak{f}) = \frac{1}{4\pi}\int_{\Gamma}dx_1^{\mu} \int_{\Gamma_\mathfrak{f}} dx_2^{\nu} \, \epsilon_{\mu\nu\rho} \frac{(x_1-x_2)^{\rho}}{|x_1-x_2|^3}\,,
\end{equation}
which is a topologically invariant quantity. Sending the deformation parameter $\delta$ to zero does not affect the integral, so that one can define the framing 
\begin{equation}
\mathfrak{f}\equiv \lim_{\delta\to 0} \chi(\Gamma,\Gamma_\mathfrak{f})\in\mathbb{Z}
\end{equation}
as the linking number between $\Gamma$ and $\Gamma_\mathfrak{f}$. More details can be found in appendix \ref{app:calugareanu}.

Considering higher-loop corrections one can check that diagrams containing at least one collapsible gauge propagator are framing-dependent \cite{Alvarez:1991sx}. As a result, framing contributions exponentiate. For a Wilson loop in the fundamental representation of the $SU(N)$ gauge group, this leads to
\begin{equation}
\label{framingCS}
    \langle W_\text{CS} \rangle_{\mathfrak{f}} = e^{\frac{i\pi N}{k}\mathfrak{f}}  \langle W_\text{CS} \rangle_{\mathfrak{f}=0}\,.
\end{equation}

Point-splitting is usually applied as a regularization method. As such, it would be unnecessary in this context. However, it provides a consistent prescription to remove the topological anomaly and obtain a knot polynomial \cite{Witten:1988hf}. As for the partition function, the scheme dependence appearing in \eqref{framingCS} is harmless, as it changes in a controlled way under a change of scheme, {\it i.e.} a change of $\mathfrak{f}$. 


\subsection{ABJ(M) theory} 

Chern-Simons-matter theories, such as ABJ(M) theory, are no longer topological, still their Wilson loop expectation values are sensitive to framing. This phenomenon is especially evident when computing them via supersymmetric localization \cite{Kapustin:2009kz,Marino:2009jd}, which selects the non-trivial framing $\mathfrak{f}=1$ \cite{Kapustin:2009kz,Drukker:2010nc}. 

As reviewed above, since the $1/6$ BPS bosonic and the $1/2$ BPS fermionic Wilson loops differ by a BRST-exact term, the identity in \eqref{eq:W12f1} holds, which has to be understood at framing $\mathfrak{f}=1$. How this identity generalizes away from this value can be investigated, for instance, by performing a perturbative evaluation of the two sides at generic framing.

Initially, perturbative results for both sides of \eqref{eq:W12f1} were obtained in ordinary perturbation theory, which corresponds to $\mathfrak{f}=0$ \cite{Bianchi:2013rma, Bianchi:2013zda, Griguolo:2013sma}. A comparison with the prediction from the matrix model \eqref{eq:MMresult} allowed to identify framing phases, up to two loops in the coupling constant, as
\begin{align}\label{eq:MMrelations}
    & \langle W^{\text{bos}} \rangle_{\mathfrak{f}=1} = e^{\frac{i\pi N_1}{k} } \langle 
    W^{\text{bos}}\rangle_{\mathfrak{f}=0}
    \,,  \qquad
     \langle \hat{W}^{\text{bos}} \rangle_{\mathfrak{f}=1} = e^{-\frac{i\pi N_2}{k} } \langle 
    \hat{W}^{\text{bos}}\rangle_{\mathfrak{f}=0}\, ,\\
     & \langle W_{1/2}\rangle_{\mathfrak{f}=1} = e^{\frac{i\pi(N_1-N_2)}{k}}\langle W_{1/2} \rangle_{\mathfrak{f}=0} \, .
    \nonumber 
\end{align}

At generic framing, an explicit perturbative computation was carried out in \cite{Bianchi:2016yzj} for the bosonic 1/6 BPS operator $W^\text{bos}$. It led to
\begin{equation}\label{eq:1/6BPSframingrelation}
    \langle W^{\text{bos}} \rangle_\mathfrak{f} = e^{\frac{i\pi N_1}{k}\mathfrak{f}- \frac{i\pi^3}{2k^3}N_1N_2^2\mathfrak{f} } \langle 
    W^{\text{bos}}\rangle_{\mathfrak{f}=0}\,,
\end{equation}  
with an analogous expression for $\hat W^{\text{bos}}$. Such a result agrees with the matrix model prediction for $\mathfrak{f}=1$ and exhibits a non-trivial correction to the framing phase arising at three loops.\footnote{Stopping the calculation at a given order in loops, the highest order framing contribution trivially exponentiates. However, it is not clear that this exponentiation persists when we push the calculation beyond that order.} This correction is entirely due to matter and in fact it is not present in the pure Chern-Simons case \eqref{framingCS}. 

A direct perturbative computation of the 1/2 BPS Wilson loop at generic framing is instead still missing, and the generalization of the identity in the second line of \eqref{eq:MMrelations} at generic framing is not known. Filling up this gap is the subject of section \ref{sec:circWLperturbative}.

Note that, while in pure Chern-Simons theories framing is simply a scheme dependence effect, in Chern-Simons-matter theories it may acquire a physical meaning. For example, in ABJ(M) theory it enters an exact prescription for computing the Bremsstrahlung function $B_{1/2}$ of the 1/2 BPS Wilson loop (see for instance \cite{Penati:2021tfj} for a self-contained review). From the general identity \cite{Bianchi:2014laa,Bianchi:2017ozk}
\begin{equation}
B_{1/2} = \frac{1}{4\pi^2} \partial_\nu \log{|\langle W^{(\nu)}_{1/6} \rangle| \Big|_{\nu=1}} \, ,
\end{equation}
where $W^{(\nu)}_{1/6}$ is the fermionic Wilson loop defined on a latitude (see eq. \eqref{eq:latitudecouplings}), exploiting the quantum cohomological equivalence for latitude operators at framing $\nu$ \cite{Bianchi:2014laa} and relying on the general structure of the bosonic Wilson loop expectation value, $\langle W^{\rm bos} \rangle_{\mathfrak{f}=1} = e^{i\Phi_B(N_1/k)} |\langle W^{\rm bos} \rangle_{\mathfrak{f}=1}|$ (where $\Phi_B$ is the $\mathfrak{f}=1$ phase), one eventually finds
\begin{equation}
    B_{1/2} = \frac{1}{8\pi} \tan{\Phi_B} \, .
\end{equation}
Therefore, the framing $\mathfrak{f}=1$ phase of the bosonic Wilson loop determines exactly the Bremsstrahlung function for the fermionic one.


\subsection{The cohomological anomaly}
\label{sec:cohomologicalanomaly}

Explicit two-loop results found in \cite{Bianchi:2013zda,Bianchi:2013rma,Griguolo:2013sma} show that $\langle W_{1/2} \rangle_{\mathfrak{f}=0}  \neq \langle W_{1/6} \rangle_{\mathfrak{f}=0}$, contrary to the na\"ive expectation based on their ${\mathcal Q}$-equivalence \eqref{eq:cohomo} at the classical level. This means that in ordinary perturbation theory -- corresponding to $\mathfrak{f}=0$ -- the cohomological equivalence between bosonic and fermionic loops is lost or, in other words, that the perturbative vacuum is not ${\cal Q}$-invariant, as it easily follows from taking the expectation value of \eqref{eq:cohomo}.

In the absence of a first-principle derivation of such a {\em cohomological anomaly}, we provide here a qualitative argument which might hint at the mechanism behind its emergence. 

The 1/6 BPS and the 1/2 BPS Wilson loops are superconformal invariant quantities, with superconformal groups SU$(1,1|1)$ and SU$(1,1|3)$, respectively. However, in perturbation theory the regularization required to tame short distance singularities breaks scale invariance at intermediate steps of the computation. This means that the stress-energy tensor acquires a non-vanishing trace, which in dimensional regularization, with $d = 3 - 2\epsilon$, is expected to be proportional to $\epsilon$. On the other hand, supersymmetry requires it to be part of a supermultiplet which contains a $\gamma$-trace for the supersymmetry currents. Consequently, the conservation of the supercharge ${\mathcal Q}$ is expected to be broken by $\epsilon$-terms as well. Explicit perturbative calculations done in dimensional regularization have shown that for circular Wilson loops the Ward identity associated with ${\mathcal Q}$-conservation acquires indeed an $\epsilon$-evanescent term \cite{Griguolo:2013sma}. We briefly recap this result.

It is convenient to consider the difference between the connections of the two loops \eqref{eq:bosonicWL1} and \eqref{eq:fermioniWL1}
\begin{equation}
\label{LminusL0}
{\cal L}-{\cal L}_0 =
\begin{pmatrix}
        -\frac{4\pi i}{k}|\dot x| C_2 \bar C^2 & -i\sqrt{\frac{2\pi}{k}}|\dot x| \eta_I \bar \psi^I \\ -i\sqrt{\frac{2\pi}{k}}|\dot x|  \psi_I \bar\eta^I & -\frac{4\pi i}{k}|\dot x| \bar C^2 C_2 
    \end{pmatrix},
\end{equation}
and call ${\cal L}_\textrm{F}$ the off-diagonal part of this expression. The diagonal terms are due to the different scalar couplings of the two operators. Following \cite{Drukker:2009hy}, the gauge function
\begin{equation}
\label{eq:lambdacircle}
    \Lambda = i\sqrt{\frac{\pi}{2k}} e^{i\tau/2}\begin{pmatrix}
        0 &&  C_2 \\
        \bar{C}^2 && 0
    \end{pmatrix}
\end{equation}
and the supercharge $\mathcal{Q}=(Q_{12+}+iS_{12+})+(Q_{34+}-iS_{34+})$ satisfying $\mathcal{Q} \Lambda = {\cal L}_\textrm{F}$ are used to derive the Ward identity 
\begin{equation}
\label{eq:Wardidcircle}
    \langle \mathcal{L}_\textrm{F}(\tau_1)\mathcal{L}_\textrm{F}(\tau_2)\rangle = 4\left[\langle\Lambda(\tau_1)D_{\tau_2}\left(e^{-i\tau_2}\Lambda(\tau_2)\right)\rangle - \langle D_{\tau_1}\left(e^{-i\tau_1}\Lambda(\tau_1)\right)\rangle\right]\,.
\end{equation}

At lowest order in perturbation theory, \eqref{eq:Wardidcircle} translates into a differential relation between the tree-level fermion and scalar propagators attached to the contour. However, when we use dimensional regularization with dimensional reduction (DRED), such a relation is broken by an evanescent term as follows
\bea
\label{eq:anomalousWardid}
    \langle (\eta\bar\psi)_1(\psi\bar\eta)_2\rangle  &= & \partial_{\tau_2} \left( e^{i\frac{\tau_{12}}{2}}\langle C_2(\tau_1)\bar C^2(\tau_2)\rangle \right) - \partial_{\tau_1} \left( e^{-i\frac{\tau_{12}}{2}}\langle C_2(\tau_1) \bar C^{2}(\tau_2)\rangle \right)  \cr
    &&-\epsilon \frac{\Gamma\left(\frac{1}{2}-\epsilon\right)}{4^{1-\epsilon}\pi^{\frac{3}{2}-\epsilon}}\frac{\left[\sin^2\frac{\tau_{12}}{2}\right]^{\frac{1}{2}+\epsilon}}{\sin \frac{\tau_{12}}{2}}\,.
\eea
These $\epsilon$-terms are harmless for the symmetry of the theory: they disappear when the regularization is removed, thus restoring superconformal invariance at the quantum level and do not affect identity \eqref{eq:cohomo} at lowest order (their loop integral vanishes). Nonetheless, they do leave a finite imprint in $1/\epsilon$-divergent diagrams, starting at two loops. This effect necessarily concurs in determining a discrepancy between the expectation values of $W_{1/2}$ and $W_{1/6}$ computed perturbatively (though it is not the only anomalous contribution at two loops). 

The cohomological anomaly for these Wilson loops can be canceled by suitably framing them, as is done in pure Chern-Simons theories to remove the topological anomaly. Up to two loops, this can be intuitively understood from the fact that, at this order, the cohomological anomaly exponentiates -- see identities \eqref{eq:MMrelations}-\eqref{eq:1/6BPSframingrelation} --, thereby producing a phase similar to framing. The observation that cohomological equivalence is respected at framing one leads to the conclusion that $\mathfrak{f}=1$ is precisely the scheme in which the framing phase compensates for the cohomological anomaly. However, in the presence of interacting matter, exponentiation of cohomologically anomalous contributions is not expected to persist at higher orders. In fact, exponentiation is typically ensured by summing over ladder contributions, but we know that at three loops, diagrams of different topologies begin contributing \cite{Bianchi:2016yzj}. Therefore, in Chern-Simons-matter theories, the compensation mechanism between cohomological anomaly and framing is more involved. A deeper understanding can be obtained by exploring the possibility of performing exhaustive perturbative calculations at non-vanishing framing, as we do in section \ref{sec:circWLperturbative}.

The quantum breaking of the cohomological equivalence resembles the breaking of invariance under inversions from the line to the circle discussed in \cite{Drukker:2000rr} for the 1/2 BPS circular Wilson loop in ${\cal N}=4$ super Yang-Mills: while the 1/2 BPS operator defined on an infinite straight line has expectation value equal to one, when defined on a circle its expectation value is no longer trivial and is entirely determined by a conformal anomaly. In the same spirit, we can state that in ABJ(M) the difference $\langle W_{1/2} \rangle - \langle W_{1/6} \rangle$ on the circle is entirely due to a cohomological anomaly. 

To support this interpretation, we observe that in the case of operators defined on the line, the off-diagonal structure of the superconnection is much simpler \cite{Drukker:2009hy}, namely
\begin{equation}
    {\cal L}_\textrm{F} = \sqrt{\frac{2\pi}{k}}\begin{pmatrix}
        0 && \eta\bar\psi_+^1 \\
        \psi_1^+ \bar\eta && 0
    \end{pmatrix}\,,
\end{equation}
with $\eta,\bar\eta$ satisfying the condition $\eta\bar\eta=2i$.
The analogue of \eqref{eq:lambdacircle} for the operator supported along the line is 
\begin{equation}
    \Lambda = \sqrt{\frac{\pi}{2k}} \begin{pmatrix}
        0 && -\eta C_2 \\
        \bar\eta \bar{C}^2 && 0
    \end{pmatrix},
\end{equation}
and the supercharge such that $\mathcal{Q}\Lambda={\cal L}_\textrm{F}$ is $\cQ=Q_{12}^+ + Q_{34+}$. In this case, the appropriate version of the Ward identity \eqref{eq:Wardidcircle} is
\begin{equation}
    \langle {\cal L}_\textrm{F}(\tau_1){\cal L}_\textrm{F}(\tau_2)\rangle = 2 \left[\Lambda(\tau_1)D_0\Lambda(\tau_2)-D_0\Lambda(\tau_1)\Lambda(\tau_2)\right]\,,
\end{equation}
which, at lowest order in perturbation theory and using DRED, translates into
\begin{equation}
    \langle \eta\bar\psi_+^1(\tau_1)\psi_1^+(\tau_2)\bar\eta\rangle = -i (\partial_{\tau_2}-\partial_{\tau_1})\langle C_2(\tau_1)\bar{C}^2(\tau_2)\rangle\,.
\end{equation}
In contrast with \eqref{eq:anomalousWardid}, the relation above is not broken by evanescent terms. This is an indication that for the line the mismatch between the vacuum expectation value of $W_{1/2}$ and $W_{1/6}$ should not occur. Indeed, for the line both operators have  expectation value equal to one.

\section{Perturbative analysis}
\label{sec:circWLperturbative}

The perturbative evaluation of the 1/2 BPS Wilson loop in ABJ(M) theory has been performed up to two loops at zero framing in \cite{Bianchi:2013zda,Bianchi:2013rma,Griguolo:2013sma}. It is a highly non-trivial computation due to the contribution of internal vertices and the fermionic couplings appearing in $W_{1/2}$. This is to be contrasted with the simplicity of the equivalent computation in ${\cal N}=4$ super Yang-Mills in four dimensions \cite{erickson}, in which only ladder diagrams contribute, directly exponentiating (in the planar limit) to the expected matrix model result.

As we have motivated above, a direct comparison with the matrix model result \eqref{eq:MMresult} requires however to consider framing $\mathfrak{f}=1$, which is what we do in the following. In fact, we keep framing completely generic at all steps of the computation and only at the end we specialize to $\mathfrak{f}=1$. To regularize divergences we employ dimensional regularization in $d=3-2\epsilon$ dimensions. We adopt the conventions and Feynman rules used in \cite{Bianchi:2013rma}, reported in appendix~\ref{app:abjm} for the reader's convenience. 

The main difficulty in computing $\vev{W_{1/2}}$ at generic framing comes from the fermionic diagrams, whose framing-dependent contributions have to be isolated. To this scope, we focus on the fermionic sector of the superconnection \eqref{eq:fermioniWL1}
\begin{equation}
\label{eq:Lf}
    {\cal L}_\textrm{F} = \begin{pmatrix}
        0 & -i\sqrt{\frac{2\pi}{k}} |\dot x| \eta_I \bar\psi^I \\  -i\sqrt{\frac{2\pi}{k}} |\dot x| \psi_I \bar\eta^I & 0
    \end{pmatrix} \equiv \begin{pmatrix}
        0 && \bar{f} \\
        f && 0 
    \end{pmatrix},
\end{equation}
and start by noting that whenever a free fermionic propagator connects two endpoints lying on the Wilson loop contour, the combination $\eta_1 \gamma^{\mu} \bar\eta_2$ appears (with $\eta_i\equiv \eta(x_i)$ and $\bar\eta_i\equiv \bar\eta(x_i)$). Using \eqref{eq:roadmapcouplingswithphases}, this reads
\begin{equation}
    \eta_1 \gamma^{\mu} \bar\eta_2 = 4i e^{i\frac{\tau_{12}}{2}} \left( \bar{s}\,\Pi_+(x_1)
     \,\gamma^\mu\, \Pi_+ (x_2) \,s\right)\,,
\end{equation}
with $\tau_{12}\equiv \tau_1-\tau_2$. Manipulations of products of $\gamma$-matrices allow to recast it as
\begin{equation}
\label{eq:bilinear}
    \eta_1 \gamma^{\mu}\bar\eta_2 = i e^{i\frac{\tau_{12}}{2}}\left[G^\mu + F^\mu \right]\,,
\end{equation}
where
\begin{equation}
\begin{split}\label{eq:G}
    G^{\mu}\equiv &\frac{\dot{x}_1^\mu}{|\dot{x}_1|}+\frac{\dot{x}_2^\mu}{|\dot{x}_2|} - i\epsilon^{\mu\nu 3}\left(\frac{\dot{x}_{1\nu}}{|\dot{x}_1|}-\frac{\dot{x}_{2\nu}}{|\dot{x}_2|}\right)+ \frac{\dot{x}_1^3 \dot{x}_2^\mu + \dot{x}_2^3 \dot{x}_1^\mu}{|\dot{x}_1| |\dot{x}_2|} + \delta^\mu_3 \left(1-\frac{\dot{x}_1\cdot\dot{x}_2}{|\dot{x}_1| |\dot{x}_2|}\right) 
\end{split}
\end{equation}
and
\begin{equation}
\begin{split}\label{eq:F}
    F^{\mu} \equiv &- i \epsilon^{\mu\nu\rho} \frac{ \dot{x}_{1\nu} \dot{x}_{2\rho}}{|\dot{x}_1|\,|\dot{x}_2|}\,.
\end{split}
\end{equation}
Recalling the definition of framing in eq. \eqref{eq:linking}, it is easy to see that framing dependence arises from the $F^\mu$ term, due to its $\epsilon^{\mu\nu\rho}$ structure. Note, in particular, that planar contours have vanishing $F^\mu$ and only receive contributions from $G^\mu$. This fact is going to be used repeatedly in the analysis.

The bilinear \eqref{eq:bilinear} is part of an effective fermion propagator for the quantity $f^{\ i}_{\hat j}$ (with $i=1,\ldots, N_1$ and $\hat j=1,\ldots, N_2$ gauge theory indices) defined in \eqref{eq:Lf}, which  we write as~\cite{Griguolo:2013sma} 
\begin{align}
\label{eq:effectivefermionpropagator2}
    \begin{split}
        \langle  \bar f^{\ \hat i}_j (x_1) f^{\ k}_{\hat l} (x_2) \rangle &=-\frac{2\pi}{k} \delta_j^k \delta_{\hat l}^{\hat i} \left( G_{12} + F_{12}\right) \, ,
    \end{split}
\end{align}
where the $G_{12}$ and $F_{12}$ functions are defined as 
\begin{align}\label{eq:GandF}
    & G_{12}\equiv\frac{1}{\epsilon} \partial_{\tau_1}\partial_{\tau_2} G(\tau_1-\tau_2)- \epsilon G(\tau_1-\tau_2)\,,\qquad 
    G(\tau_1-\tau_2) \equiv \frac{\Gamma(\frac{1}{2}-\epsilon)}{4^{1-\epsilon}\pi^{\frac{3}{2}-\epsilon}} \frac{\left[\sin^{2}\frac{\tau_1 - \tau_2}{2}\right]^{\frac{1}{2}+\epsilon}}{\sin\frac{\tau_1 -\tau_2}{2}}\,, \nonumber\\&
    F_{12}\equiv\frac{e^{i\frac{\tau_{12}}{2}}}{4\pi i}\epsilon_{\mu\nu\rho} \dot x_1^\mu \dot x_2^\nu \frac{(x_1 -x_2)^\rho}{|x_1-x_2|^{3}}\,,
\end{align}

The $G_{12}$ function comes from the framing-independent terms in \eqref{eq:G}, specialized to the circular contour \eqref{eq:maximalcircle}. It is evaluated in dimensional regularization at $d=3-2\epsilon$, since intermediate steps of the calculation require a regulator. This function is manifestly real and antisymmetric in its argument. The term in $G_{12}$ proportional to $\epsilon$ is zero for $\epsilon\to 0$, but it contributes at two loops when multiplied by singular contributions going like $1/\epsilon$. This is an explicit realization of the general mechanism discussed in section \ref{sec:cohomologicalanomaly}. 

The $F_{12}$ function comes instead from \eqref{eq:F}. We will see that it corresponds to a framing-dependent term in the fermionic propagator. For this reason we keep it generic, not specializing it to the circle, where it would vanish. This term possesses an imaginary component which is symmetric under the exchange $\tau_1\leftrightarrow\tau_2$, and a real part which is antisymmetric. We will ascertain that the former sources a framing-dependent term in the fermionic propagator, whereas the latter yields vanishing contributions in all the diagrams we consider. The analysis of the $F_{12}$ function will not require regularization, so in its definition we have set $\epsilon=0$.

A comment about the choice of fermionic couplings is now in order. The expression for $F_{12}$ in \eqref{eq:GandF} rests on the definition \eqref{eq:roadmapcouplingswithphases}, which ensures supersymmetry for the Wilson loop on the circle. The framing procedure deforms the contour away from the circle, thereby breaking supersymmetry. This seems counterproductive, as framing with Hopf fiber circles was advocated as a supersymmetry preserving regularization procedure for localization, ultimately motivating why its predictions hold at framing 1 \cite{Kapustin:2009kz}. Our working assumption is that the effects of the deformation we are considering are negligible in our perturbative calculation. In fact, the breaking is reflected in potential terms of order $\delta$ -- the framing deformation parameter defined in \eqref{eq:framedcontour} -- in the numerators of the various contour integrals. Such pieces are expected to vanish in the $\delta\to 0$ limit, because framing effects already require delicate cancellations of powers of $\delta$ between numerators and denominators of integrals. We explicitly verified this fact at one loop, where a simple variation of the Gauss linking integral appears. Our hypothesis is that the same mechanism holds at two loops for more complicated integrals as well. We also corroborated the robustness of our procedure, by using the fermionic couplings of \cite{Cardinali:2012ru}, which enforce supersymmetry on generic contours on a two-sphere. This choice produces the same results, however the analogous version of \eqref{eq:GandF} is bulkier and it encumbers explicit calculations, especially at two loops. Therefore we chose to stick to \eqref{eq:GandF}.


\subsection{First order in perturbation theory}
\label{sec:firstorder}

At first order in perturbation theory, the only diagrams contributing to $\vev{W_{1/2}}$ are {\it i)}~the single gauge field exchange and {\it ii)} the single fermion exchange. There is no analogue for the scalars, which enter the Wilson loop in pairs and then start contributing at two loops.

The one-loop gauge field exchange at generic framing has been computed in pure Chern-Simons theory in \cite{Guadagnini:1989am}. For an unnormalized Wilson loop in the fundamental representation of $U(N)$ it is given by
\bea
\vev{W_\textrm{CS}}_\mathfrak{f}= \frac{i \pi}{k} N^2 \, \mathfrak{f}.
\eea
This result applies immediately to $\vev{W^\textrm{bos}}$ in ABJ(M), as well as to $\vev{\hat W^\textrm{bos}}$, modulo a sign flip due to the opposite sign of the Chern-Simons level in the second node of the quiver. The unnormalized result for the 1/2 BPS operator is then the sum of these two contributions (we denote gauge fields with wavy lines throughout the paper)
\begin{equation}
\label{eq:1loopGauge}
    \vcenter{\hbox{\includegraphics[width=0.1\textwidth]{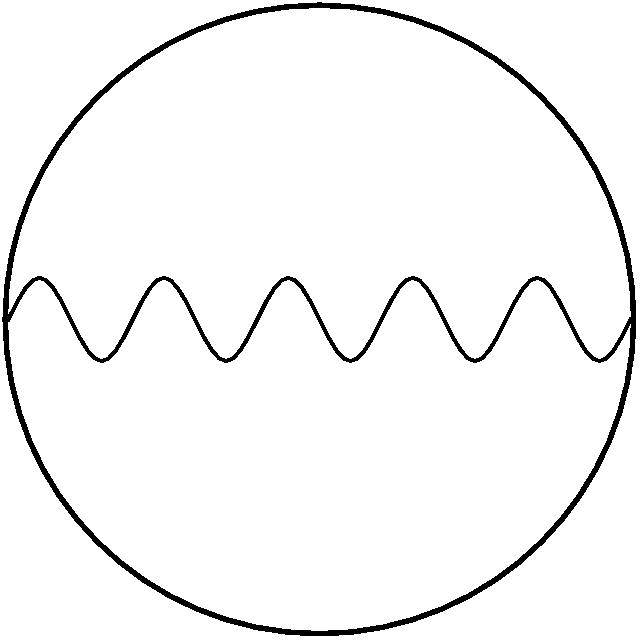}}} \quad = \quad  \frac{i \pi}{k} \, (N_1^2-N_2^2) \, \mathfrak{f}\,.
\end{equation}
Normalizing as in \eqref{eq:fermioniWL1} by $\text{STr}(\mathcal{T})=N_1+N_2$, we obtain 
\bea
\label{eq:1loopGauge1}
\vev{W_{1/2}}_\mathfrak{f}{}\Big|_{\; i)} =  \frac{i\pi}{k} \, (N_1 - N_2) \, \mathfrak{f}.
\eea
For $\mathfrak{f}=1$, this matches the result expected from the matrix model at this order, see the second term in \eqref{eq:MMresult}. 

The contribution from a single fermion exchange is given by (fermions are denoted by solid lines)
\begin{equation}\label{eq:cF}
    \vcenter{\hbox{\includegraphics[width=0.1\textwidth]{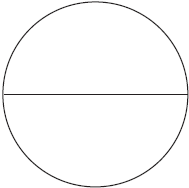}}}  \quad = -\int d\tau_{1>2}\left[ \langle\bar f_1 f_2\rangle +\langle 
        f_1\bar f_2 \rangle \right]\,,
\end{equation}  
where $f_{1,2} \equiv f(x_{1,2})$ (and similarly for the barred quantities) are the fermionic off-diagonal elements in \eqref{eq:Lf}. We focus on the first contribution, the second one being simply its complex conjugate. Inserting the fermionic propagator \eqref{eq:effectivefermionpropagator2}, it reads
\begin{equation}\label{eq:ff}
   \int d\tau_{1>2} \, \langle\bar f_1 f_2\rangle = -\frac{2\pi}{k} N_1N_2  \int d\tau_{1>2} \ (G_{12} + F_{12} )\,,
\end{equation}
with $G_{12}$ and $F_{12}$ given in \eqref{eq:GandF}.

To investigate what happens to $G_{12}$ at non-trivial framing, we use point-splitting regularization and take $x_2^\mu$ on the framed contour \eqref{x2contour}. As originally done in \cite{calugareanu1959integrale}, to single out possible framing-dependent integrals we expand the resulting expression in powers of $\delta$. At order $\mathcal O(\delta^0)$, we recover the zero-framing integral, which was shown to evaluate to zero \cite{Bianchi:2013zda,Bianchi:2013rma,Griguolo:2013sma}. Higher order terms in the expansion disappear in the $\delta\to0$ limit. 

On the other hand, the term with $F_{12}$ could yield, in principle, a framing-dependent contribution. 
The corresponding integral is finite, thus we can set the regularization parameter $\epsilon$ to zero. Apart from the extra phase $e^{i\frac{\tau_{12}}{2}}$, the rest of the integrand is exactly the one appearing in the Gauss linking number \eqref{eq:linking}. As proven in appendix \ref{app:calugareanu}, 
for any analytic function $h(\tau)$ multiplying the Gauss integrand, the integral results into $h(0)$ times the Gauss linking number. Since in the case at hand, $h(\tau)=e^{i\frac{\tau}{2}}$, with $h(0)=1$, the result is nothing but the Gauss linking number. 
Therefore,
\begin{equation}
\label{eq:1loopresultF}
    \int d\tau_{1>2} \, \langle\bar f_1 f_2\rangle = \, \frac{2\pi i}{k} \,N_1N_2 \,  \mathfrak{f} \,.
\end{equation}

Adding the complex conjugate contribution in \eqref{eq:cF}, one finds that the single fermion exchange at generic framing vanishes identically
\begin{equation}\label{eq:framinfResult1}
    \vcenter{\hbox{\includegraphics[width=0.1\textwidth]{img/1loopfe.png}}}  \quad = - \frac{2\pi i}{k} \,N_1N_2 \,  \mathfrak{f}+ \frac{2\pi i}{k} \,N_1N_2 \,  \mathfrak{f}= 0.
\end{equation}
In particular, it vanishes at $\mathfrak{f}=1$, consistently with the matrix model prediction \eqref{eq:MMresult}, which at this order is entirely given  by the gauge field exchange \eqref{eq:1loopGauge1}.

We conclude this section with a non-trivial check of the method that led to \eqref{eq:framinfResult1}, considering the $1/6$ BPS fermionic latitude Wilson loop reviewed in section \ref{sec:WLsABJM}. In this case cohomological equivalence holds at framing $\mathfrak{f}=\nu$, with the expectation value of bosonic operators being captured by a deformed matrix model \cite{Bianchi:2018bke} containing an explicit dependence on the latitude parameter $\nu$. This is an interesting example in which framing get analytically continued to a non-integer value. The one-loop contribution from fermionic diagrams reads in this case
\begin{equation}
    \vcenter{\hbox{\includegraphics[width=0.1\textwidth]{img/1loopfe.png}}}  \quad = -\int d\tau_{1>2}\left[ e^{-\frac{i\pi\nu}{2}}\langle\bar f_1 f_2\rangle - e^{\frac{i\pi\nu}{2}}\langle 
        f_1\bar f_2 \rangle \right]   \,,
\end{equation}  
where $f$ and $\bar f$ are now the fermionic off-diagonal elements with the couplings in \eqref{eq:latitudecouplings}. The result at framing zero is known to be \cite{Bianchi:2014laa}
\begin{equation}
\left.\vcenter{\hbox{\includegraphics[width=0.1\textwidth]{img/1loopfe.png}}}\,\,\right|_{\mathfrak{f}=0} = \frac{2\pi i}{k} \,N_1N_2 \,  \nu \cos\frac{\pi\nu}{2}.
\end{equation}
The framing part turns out to be the same as above even for the latitude contour \eqref{eq:latitudecontour}, thus giving
\begin{align}\label{eq:latitudeResult}
    \begin{split}
\vcenter{\hbox{\includegraphics[width=0.1\textwidth]{img/1loopfe.png}}}  \quad &=\quad     \left.\vcenter{\hbox{\includegraphics[width=0.1\textwidth]{img/1loopfe.png}}}\,\,\right|_{\mathfrak{f}=0}
    -  \frac{i\pi}{k} e^{-\frac{i\pi\nu}{2}} \, N_1N_2 \, \mathfrak{f} -  \frac{i\pi}{k}  e^{\frac{i\pi\nu} {2}} \,  N_1N_2 \,  \mathfrak{f}\\
    &= \frac{2\pi i}{k} \, N_1N_2 \, (\nu-\mathfrak{f})\cos\frac{\pi\nu}{2}\,.       
    \end{split}
\end{align}
where we have used the one-loop result at trivial framing known from \cite{Bianchi:2014laa}.

This expression vanishes for $\mathfrak{f}=\nu$, which is consistent with the statement above that the matrix model computes the expectation value at framing $\nu$. For this value of framing, cohomological equivalence between fermionic and bosonic Wilson loops holds. In order for this condition to be satisfied, the total contribution of fermionic diagrams at one loop should vanish at framing $\nu$, which is indeed the case. 


\subsection{Second order in perturbation theory}

At two loops, the results for the gauge field diagrams at generic framing can be directly extracted from the pure Chern-Simons case \cite{Guadagnini:1989am}. In ABJ(M) theory there is one extra bosonic diagram with the exchange of two scalars \cite{Drukker:2008zx,Chen:2008bp,Rey:2008bh}, which however is not affected by framing, as it is simply the square of the scalar propagator. 

In summary, the bosonic diagrams at generic framing for the unnormalized $W^{\text{bos}}$ operator~\eqref{eq:Wbos} are (we indicate scalars with dotted lines)
\begin{equation}
\begin{split}
    \label{eq:bos2loop}
    \vcenter{\hbox{\includegraphics[width=0.1\textwidth]{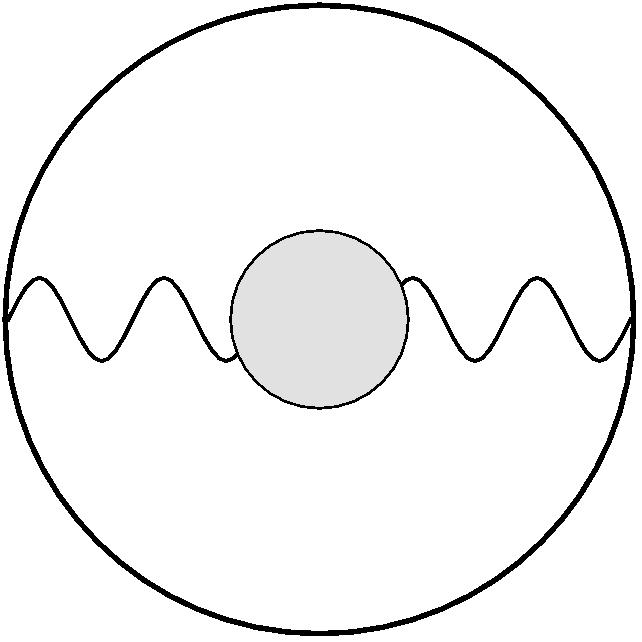}}} + \vcenter{\hbox{\includegraphics[width=0.1\textwidth]{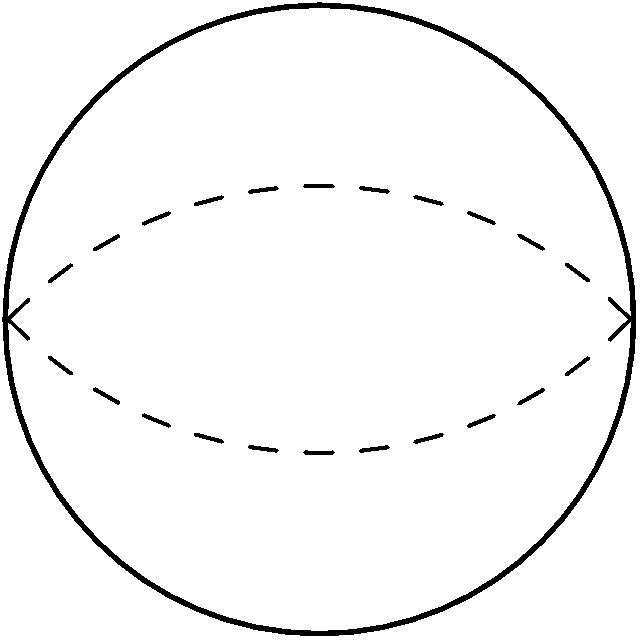}}} \quad&= \quad \frac{\pi^2}{k^2} \, N_1^2 N_2 \,, \\
    \vcenter{\hbox{\includegraphics[width=0.1\textwidth]{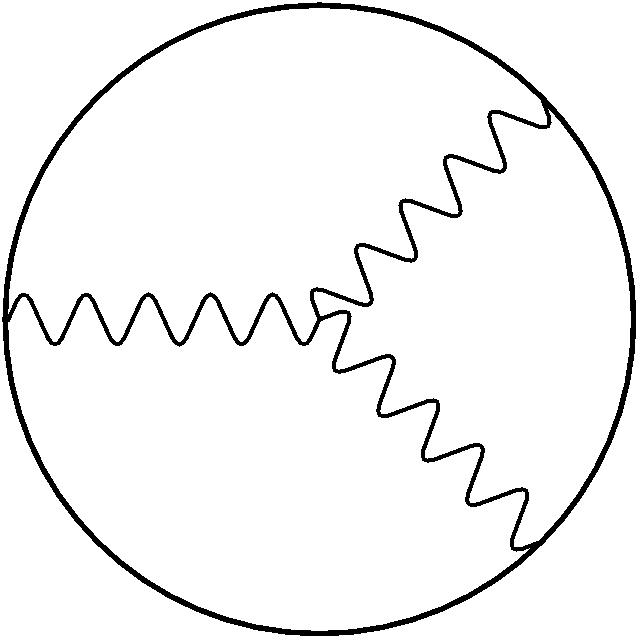}}}\quad &=\quad- \frac{\pi^2}{6k^2} \, N_1\left(N_1^2-1\right)  \,, \\
    \vcenter{\hbox{\includegraphics[width=0.1\textwidth]{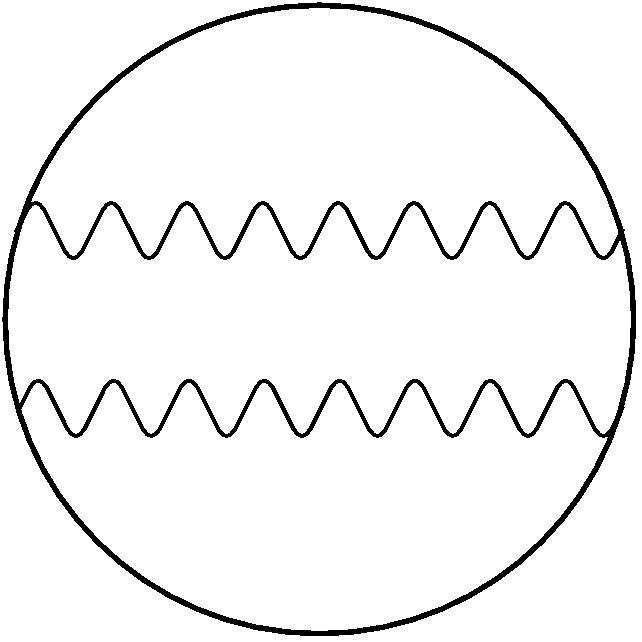}}} \quad&= \quad - \frac{\pi^2}{2k^2} \, N_1^3 \, \mathfrak{f}^2 \,.
\end{split}
\end{equation}
The corresponding results for $\hat W^{\text{bos}}$ are obtained by exchanging $N_1$ and $N_2$.

The non-trivial and novel part of the computation is instead represented by the diagrams with fermions, given by
\begin{itemize}
    \item[{\it i)}] the one-loop correction of a single-fermion exchange;
    \item[{\it ii)}] a mixed gauge-fermion exchange;
    \item[{\it iii)}] a double-fermion exchange;
    \item[{\it iv)}] a cubic gauge-fermion vertex diagram.
\end{itemize}
In what follows we compute them separately at generic framing.

The ordinary perturbative evaluation of the 1/2 BPS Wilson loop at framing zero provides a vanishing result at one loop and a real result at two loops. On the other hand, the comparison with the matrix model result at this order allows to identify the framing phase, see \eqref{eq:MMrelations}, whose expansion still provides a real two-loop result at framing one. If we reasonably assume that a calculation at generic framing should only modify the exponent of the framing phase without spoiling the general structure of identity \eqref{eq:MMrelations}, we expect a real result at two loops for any framing. Based on this argument, along the calculation we are going to neglect imaginary contributions arising from two-loop diagrams, as they will eventually cancel in the final result.


\subsubsection*{{\it i)} One-loop corrected single-fermion exchange}

The first fermionic diagram is the self-energy correction to the single fermion exchange. At zero framing this was shown to be zero in \cite{Bianchi:2013rma}. It is easy to see that turning on framing does not change this conclusion. In fact,  framing contributions may only arise from non-planar terms. On the other hand, the structure of the corrected propagator, see \eqref{eqn:onelooppropagator} in the appendix, does not allow for these terms to appear. We then obtain
\begin{equation}
\begin{split}
    \vcenter{\hbox{\includegraphics[width=0.1\textwidth]{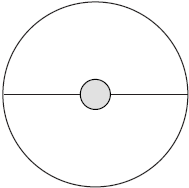}}}  \quad&= 0\,.
\end{split}
\end{equation}


\subsubsection*{{\it ii)} Mixed gauge-fermion exchange diagram}

This diagram does not appear in the zero framing analysis, since the structure of the gauge propagator (see \eqref{eqn:propagator}) makes it automatically vanish for planar contours. At non-vanishing framing, however, non-trivial contributions may arise, which we now compute. 

There are two mixed gauge-fermion exchange diagrams, corresponding to  the $A_\mu$ and $\hat{A}_\mu$ propagators. They give rise to equal contributions, apart from the color factors, which are $N_1^2N_2$ and $N_2^2N_1$, respectively. We focus on the former, which schematically reads
\begin{equation}\label{eq:mixedgauge1}
    N_1^ 2N_2 \left( \langle A_1 A_2 \rangle \langle \bar f_3 f_4 \rangle + \langle 
    A_3 A_4 \rangle \langle \bar f_1 f_2 \rangle +  \langle A_1 A_4 \rangle \langle \bar f_2 f_3 \rangle -  \langle A_2 A_3 \rangle \langle \bar f_4 f_1 \rangle \right)\,,
\end{equation}
where the minus sign in front of the last term is due to the anticommutation of fermions. Here $A_i$ stands for $A_\mu(x_i)$, and similarly for $f_i$ and $\bar f_i$.

Inserting the effective propagators \eqref{eq:effectivefermionpropagator2} in \eqref{eq:mixedgauge1}, the computation splits into a `framing squared' part, given by the product of one gauge propagator and the framing term in the fermionic one (the $F$-term in \eqref{eq:effectivefermionpropagator2}) and a `linear framing' part, given by the product of one gauge propagator and the planar sector of the fermionic propagator (the $G$-term in \eqref{eq:effectivefermionpropagator2}). We argue that only the `framing squared' part is relevant for our computation, since the other contribution turns out to be imaginary. Therefore, generalizing the definition of the framing-dependent term $F_{12}$ in \eqref{eq:GandF} to generic points
\bea
F_{ij}=\frac{e^{i\frac{\tau_{ij}}{2}}}{4\pi i}\epsilon_{\mu\nu\rho} \dot x_i^\mu \dot x_j^\nu \frac{(x_i -x_j)^\rho}{|x_i-x_j|^{3}},
\eea
we focus on the computation of
\begin{equation}
\label{eq:mixedgauge2}
    N_1^ 2N_2 \left( \langle A_1 A_2 \rangle F_{34} + \langle 
    A_3 A_4 \rangle F_{12} +  \langle A_1 A_4 \rangle F_{23} -  \langle A_2 A_3 \rangle F_{41} \right)\,.
\end{equation}
We manipulate this expression in order to write it as the product of the one-loop gauge diagram times the one-loop fermionic one, up to framing-independent terms. To this end, we sum and subtract the two cross contributions
\begin{equation}\label{eq:additional0}
N_1^ 2N_2 \left( \langle A_1 A_3 \rangle F_{24} + \langle A_2 A_4 \rangle  F_{13} \right),
\end{equation}
and an additional $2\langle A_2 A_3 \rangle F_{41}$, in order to have the same sign for all the terms in \eqref{eq:mixedgauge2}. Integrating over the four endpoints of the propagators, appropriately ordered, we are led to evaluate
\begin{align}\label{eq:additional}
    & N_1^ 2N_2 \bigg[\left( \int_{\tau_1>\tau_2}\langle A_1 A_2 \rangle \right) \left( \int_{\tau_3>\tau_4}F_{34}\right)
    \nonumber\\&~~~~~~~~~~~
    - \int_{\tau_1>\tau_2>\tau_3>\tau_4}\left( \langle A_1 A_3 \rangle F_{24} + \langle A_2 A_4 \rangle F_{13}\right)
    -2\, \int_{\tau_1>\tau_2>\tau_3>\tau_4} \langle A_2 A_3 \rangle F_{41}
    \bigg]\,.
\end{align}

First, we show that all the terms in the second line of \eqref{eq:additional} are framing-independent, thus they enter the result at zero framing and can be disregarded in the present analysis. At a first sight, their independence on framing might be inferred from the fact that they all contain at least one non-collapsible propagator, while it is believed that framing dependence only arises from ``collapsible'' propagators, {\it i.e.} propagators connecting $\tau_i$ with $\tau_{i+1}$ \cite{Alvarez:1991sx}. However, since this argument is valid only in pure Chern-Simons theory and does not necessarily hold in the presence of matter \cite{Bianchi:2016yzj}, we check framing independence explicitly via a numerical computation.

\begin{figure}[!ht]
    \centering
    \includegraphics[width=0.55\textwidth]{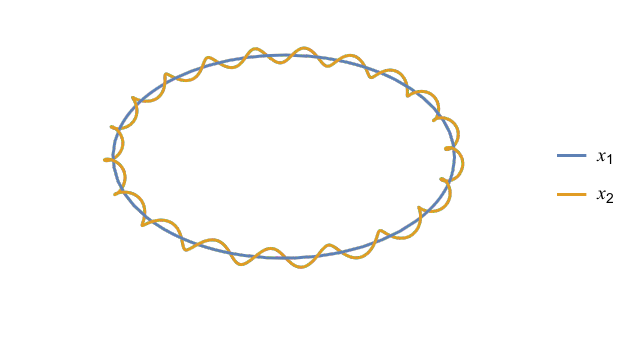}
    \caption{The contour $x_1$ is the original circle \eqref{eq:maximalcircle}, while $x_2$ is an helix with radius $\delta = 0.05$ and winding number $\mathfrak{f} = 20$.}
    \label{fig:helix}
\end{figure}

\begin{figure}[!htp]
\centering
    \includegraphics[width=0.9\textwidth]
    {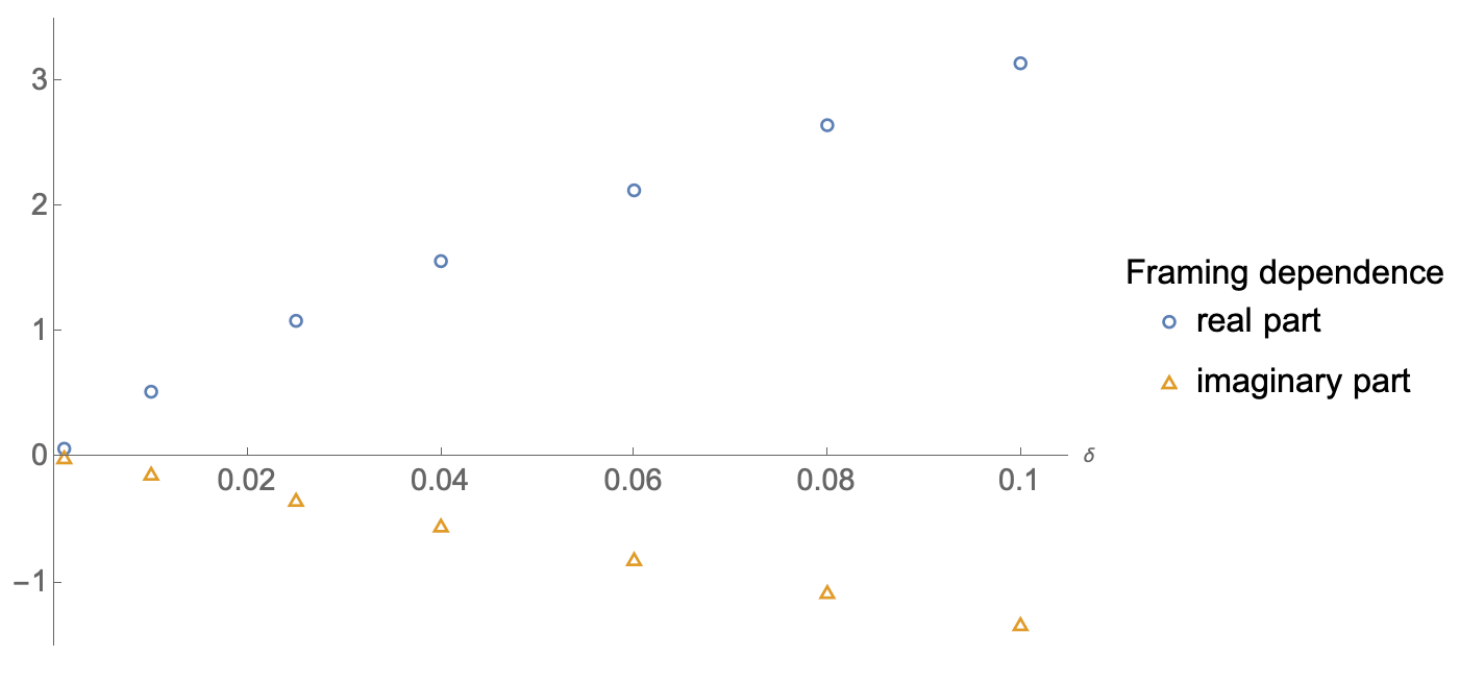}
    \caption{Real and imaginary parts of the framing dependence of the second line of \eqref{eq:additional} for $\mathfrak{f}=1$ as a function of~$\delta$. Both parts tend to zero as $\delta\to 0$, proving numerically that those expressions are in fact framing-independent.}
\label{fig:FramingFermGaugev2}
\end{figure}

We proceed as follows. Without loss of generality, we let $\tau_1$  run on the circle \eqref{eq:maximalcircle} while the three parameters $\tau_{2,3,4}$ run on small deformations of it
\begin{equation}\label{eq:2loopParametrization}
    x_{k=2,3,4}^{\mu} = (\cos\tau_k,\sin\tau_k,0) + (k-1)\, \delta \, (\cos (\mathfrak{f}\, \tau_k)\cos\tau_k, \cos (\mathfrak{f}\,\tau_k) \sin\tau_k, \sin (\mathfrak{f}\, \tau_k))\, .
\end{equation}
This consists of a toroidal helix of infinitesimal radius $\delta$ winding $\mathfrak{f}$ times around the original circular path, see figure \ref{fig:helix}. 
We evaluate numerically the framing of the second line of \eqref{eq:additional} for $\mathfrak{f}=1$ and different values of $\delta$, see figure \ref{fig:FramingFermGaugev2}. It can be clearly seen from the plot that framing disappears as $\delta$ is sent to zero, confirming our claim.

Having established this fact, we can now compute the first line of \eqref{eq:additional} by reading it as the product of the one-loop gauge diagram \eqref{eq:1loopGauge} and one-loop fermion diagram \eqref{eq:1loopresultF}. From the upper block of the superconnection we obtain a framing-dependent term which reads
\begin{equation}
\label{eq:fsquare}
    \frac{\pi^2}{k^2} \, N_1^2N_2  \, \mathfrak{f}^2\,.
\end{equation}
Adding also the real contribution from the lower block of the superconnection, the final result for the (unnormalized) mixed gauge-fermion exchange diagram is
\begin{equation}\label{eq:mixedResult}
\vcenter{\hbox{\includegraphics[width=0.08\textwidth]{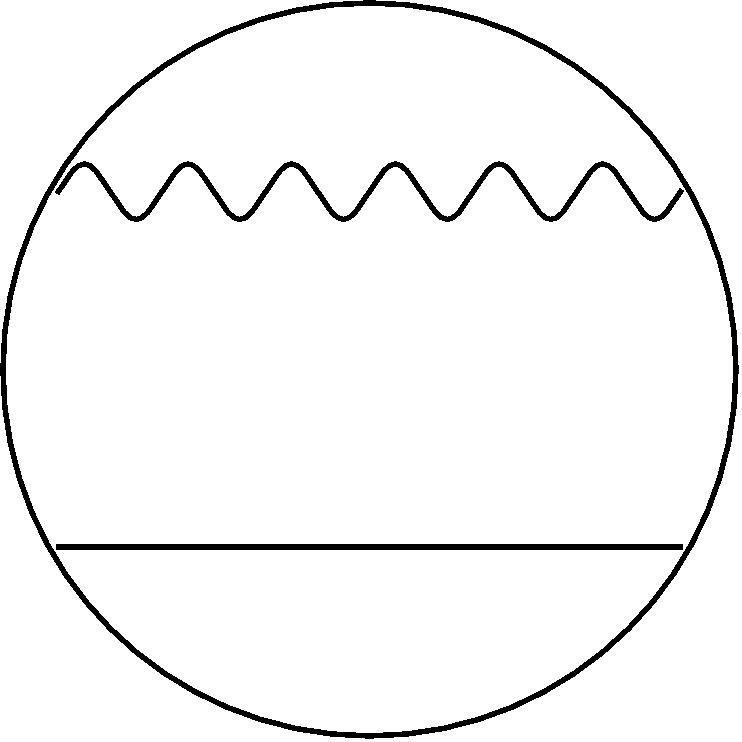}}} =  \frac{\pi^2}{k^2} \, N_1N_2(N_1+N_2) \, \mathfrak{f}^2   \,.
\end{equation}
Recalling that this diagram does not contribute at zero framing (the gauge propagator makes it vanish on planar contours), this is its entire real contribution. 


\subsubsection*{{\it iii)} Double fermion exchange diagram}

The double fermion exchange diagram comes from the following terms 
\begin{equation}
    \langle \bar f_1 f_2 \bar f_3 f_4 \rangle + \langle f_1 \bar f_2 f_3 
    \bar f_4 \rangle\,.
\end{equation}
For both one can perform contractions in two ways, producing different overall color factors
\begin{equation}\label{eq:doublefermionexchange1}
    N_1N_2^2\left(\langle \bar f_1 f_2 \rangle \langle \bar f_3 f_4 \rangle - \langle\bar f_4 f_1 \rangle \langle \bar f_2 f_3 \rangle \right) +N_1^2N_2\left( - \langle \bar f_1 f_4 \rangle \langle \bar f_3 f_2 \rangle + \langle\bar f_2 f_1 \rangle \langle \bar f_4 f_3 \rangle \right).
\end{equation}
Focusing on the contribution proportional to $N_1N_2^2$ and inserting the effective propagator \eqref{eq:effectivefermionpropagator2} we find
\bea\label{eq:doublefermioncomputation1}\nonumber
    \int_{\tau_1>\tau_2>\tau_3>\tau_4} \langle \bar f_1 f_2 \rangle \langle \bar f_3 f_4  \rangle =
    \frac{4\pi^2}{k^2}\int_{\tau_1>\tau_2>\tau_3>\tau_4} \Big[G_{12} G_{34} + G_{12}F_{34} + F_{12}G_{34} + F_{12}F_{34}\Big]\,,\\
\eea
and similarly for $\langle\bar f_4 f_1 \rangle \langle \bar f_2 f_3 \rangle$. The first term on the right hand side gives the previously known zero framing contribution. Its detailed computation can be found in \cite{Griguolo:2013sma}. Including also the $N_1^2 N_2$ color structure, the zero framing (unnormalized) result reads
\begin{equation}\label{eq:firstline}
    \left.\vcenter{\hbox{\includegraphics[width=0.1\textwidth]{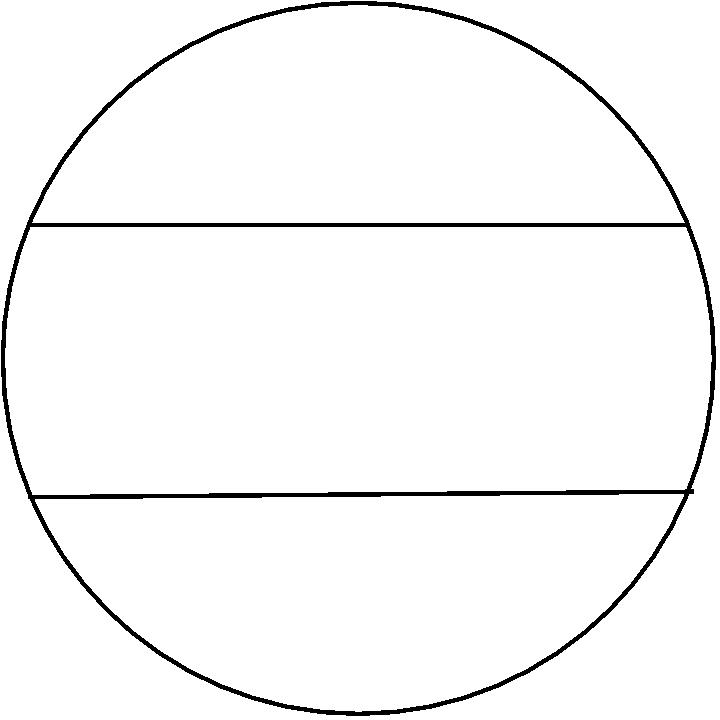}}}\,\,\right\vert_{\mathfrak{f}=0}  = \frac{3\pi^2}{2k^2}\, N_1N_2(N_1+N_2)\,.
\end{equation}
The rest of \eqref{eq:doublefermioncomputation1} contains framing-dependent integrals. To evaluate them we proceed as we did for the mixed gauge-fermion diagram.  

In the evaluation of the second and third terms, we discard order $\epsilon$ contributions, since $G$ is constant in the $\epsilon\to 0$ limit. For $\frac{1}{\epsilon}$-terms, two integrations can be trivially performed exploiting the derivatives and, once combined with the ones coming from $\langle\bar f_4 f_1 \rangle \langle \bar f_2 f_3 \rangle$, they cancel each other.

\begin{figure}[!ht]
    \centering
    \includegraphics[width=0.9\textwidth]
    {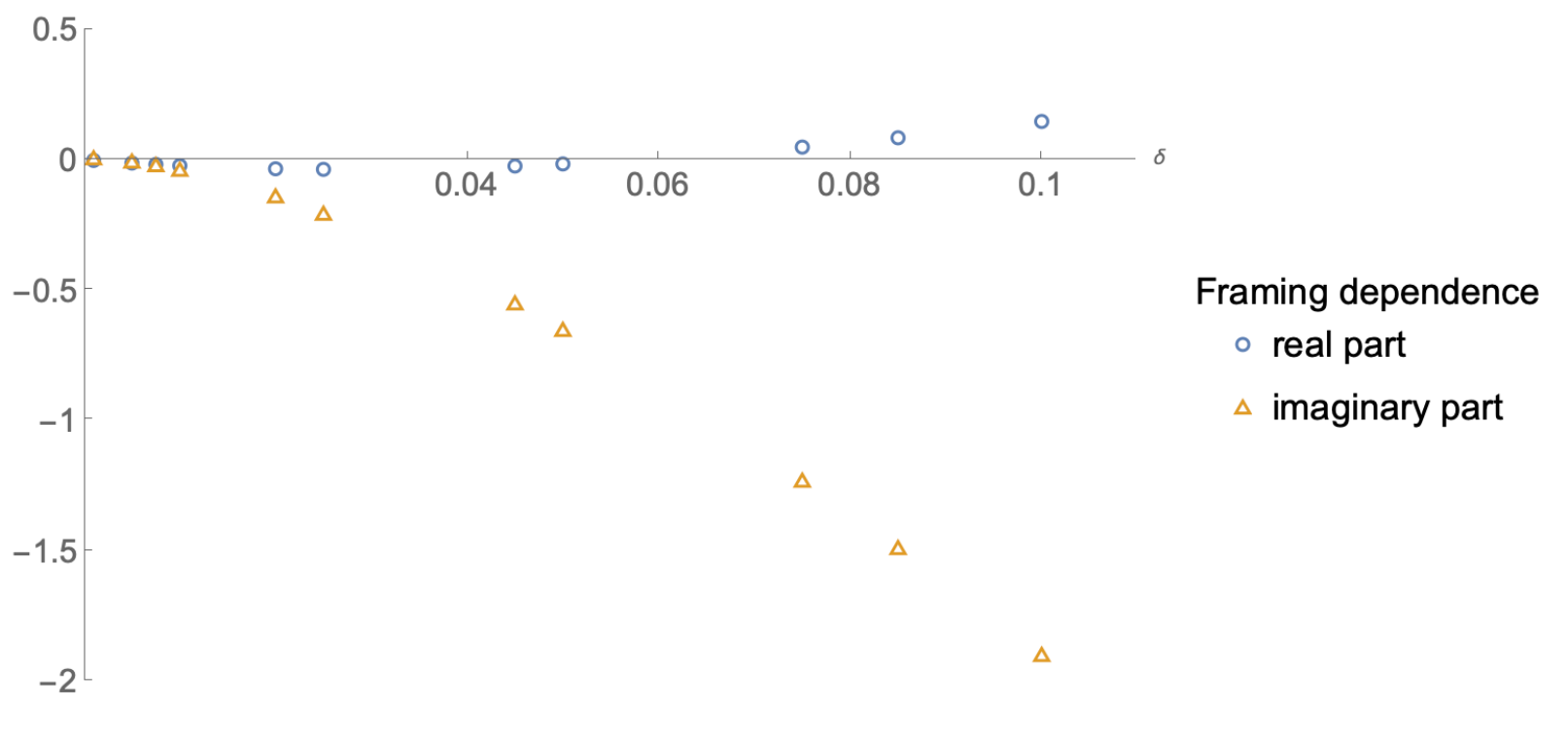}
    \caption{Real and imaginary parts of the framing dependence of   $\int (F_{13}F_{24} + 2 F_{41}F_{23})$ in \eqref{eq:doublefermionSym} for $\mathfrak{f}=1$ and different values of $\delta$. Both parts tend to zero as $\delta\to 0$, proving numerically that those terms are in fact framing-independent.}
\label{fig:doubleferm}
\end{figure}

In order to compute the last term of \eqref{eq:doublefermioncomputation1}, and the equivalent one from $\langle\bar f_4 f_1 \rangle \langle \bar f_2 f_3 \rangle$ as well, we symmetrize the integrand by adding and subtracting crossed terms. This allows to  factorize it as the square of the one-loop result plus extra terms. Precisely,
\begin{equation}\label{eq:doublefermionSym}
    \int_{\tau_1>\tau_2>\tau_3>\tau_4} \hskip -.5cm \left(F_{12}F_{34} - F_{41}F_{23} \right) = 
    \frac{1}{2} \left(\int_{\tau_1>\tau_2} \hskip -.5cm  F_{12}\right)^2 - \int_{\tau_1>\tau_2>\tau_3>\tau_4} \hskip -.5cm  \left( F_{13}F_{24} + 2F_{41}F_{23}\right)\,.
\end{equation}
Parameterizing the contour as in \eqref{eq:2loopParametrization}, one can check numerically that the two extra terms on the right hand side of \eqref{eq:doublefermionSym} are framing-independent and vanish in the $\delta \to 0$ limit, as shown in figure \ref{fig:doubleferm}.
Therefore, the only non-trivial contribution in \eqref{eq:doublefermionSym} comes from the square of the one-loop result. 

Including the contributions proportional to the $N_1^2N_2$ color factor and summing with \eqref{eq:firstline}, we find 
\begin{equation}\label{eq:doublefermionexchangeresult}
    \vcenter{\hbox{\includegraphics[width=0.1\textwidth]{img/fd6.jpg}}}  =  \frac{\pi^2}{2k^2} \, N_1N_2(N_1+N_2) \,  (3 - \mathfrak{f}^2) \,. 
\end{equation}


\subsubsection*{{\it iv)} Gauge-fermion vertex diagram}

Due to the presence of a bulk vertex, the evaluation of the gauge-fermion vertex diagram is highly non-trivial, even at framing zero \cite{Bianchi:2013zda,Bianchi:2013rma,Griguolo:2013sma}. Its  contribution can be cast in the form $(\mathcal{X}_{12;3} + \mathcal{X}_{13;2} + \mathcal{X}_{23;1} )$, where \cite{Griguolo:2013sma}
\begin{equation}\label{eq:vertexmiddlestep}
    \mathcal{X}_{12;3} \equiv \int d\tau_{1>2>3}\left[ \delta_\nu^\mu \eta_1 \gamma_{\lambda} \bar\eta_2 +\delta_\lambda^\mu \eta_1 \gamma_\nu \bar\eta_2 - \delta_{\lambda\nu} \eta_1 \gamma^\mu \bar\eta_2 - i \eta_1\bar\eta_2 \epsilon^{\mu}_{\ \nu\lambda} \right] \epsilon_{\mu\rho\sigma}\dot x_3^\rho \Gamma^{\nu\lambda\sigma}\,,
\end{equation}
with the other two terms given by a permutation of the indices inside the integrand. Here $\Gamma^{\nu\lambda\sigma}$ is the following bulk integral
\begin{equation}
    \Gamma^{\nu\lambda\sigma} = \left(\frac{\Gamma(\frac{1}{2}-\epsilon)}{4\pi^{\frac{3}{2}-\epsilon}}\right)^3 \partial_{x_1^{\nu}} \partial_{x_2^{\lambda}}\partial_{x_3^{\sigma}}\int \frac{d^{3-2\epsilon}w}{|x_{1w}|^{1-2\epsilon}|x_{2w}|^{1-2\epsilon}|x_{3w}|^{1-2\epsilon}}\,.
\end{equation}

For our purposes it is sufficient to focus on $\mathcal{X}_{12;3}$, as the other terms can be treated in a similar way. We proceed by first inserting the explicit expression \eqref{eq:bilinear} for the spinor bilinear $\eta_1\gamma \bar\eta_2$ in \eqref{eq:vertexmiddlestep}. This gives rise to a complex integrand, whose real part leads precisely to the zero framing result of \cite{Bianchi:2013zda,Bianchi:2013rma,Griguolo:2013sma}. Extra contributions due to the non-trivial framing then arise necessarily from the imaginary part. However, as already mentioned, imaginary contributions should eventually cancel at two loops, therefore we can safely discard them. In conclusion, this diagram does not contribute to framing and its real part simply coincides with the framing zero result
\begin{equation}
\label{eq:gaugefermionvertex}
    \vcenter{\hbox{\includegraphics[width=0.1\textwidth]{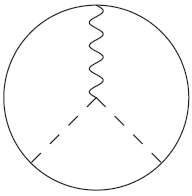}}}\,\, =- \frac{2\pi^2}{k^2} \, N_1N_2(N_1+N_2)\,. 
\end{equation}


\subsection*{Final result}

Combining the results above and normalizing them by $\textrm{STr}({\cal T})=N_1+N_2$, we obtain the complete two-loop expectation value of the 1/2 BPS fermionic Wilson loop in ABJ(M) theory, at generic framing $\mathfrak{f}$
\begin{equation}\label{eq:W12result}
    \vev{W_{1/2}}_\mathfrak{f} = 1 + \frac{i\pi}{k}(N_1-N_2) \, \mathfrak{f} - \frac{\pi^2}{6k^2}\left[ 
           \left(N_1^2 -4N_1N_2+ N_2^2-1\right) +  3(N_1-N_2)^2 \, \mathfrak{f}^2 \, \right] +\mathcal{O}\left( k^{-3}\right).
\end{equation}

This is the main result of our analysis. At  $\mathfrak{f}=0$, it correctly reproduces known perturbative results \cite{Bianchi:2013zda,Bianchi:2013rma,Griguolo:2013sma}. At $\mathfrak{f}=1$, it coincides with the localization result \eqref{eq:MMresult} and confirms the cohomological equivalence with the 1/6 BPS bosonic Wilson loop \cite{Marino:2009jd}. At this order, it explicitly exhibits exponentiation of framing
\begin{equation}\label{eq:W12result1}
    \vev{W_{1/2}}_\mathfrak{f} = e^{\frac{i\pi}{k}(N_1-N_2)\mathfrak{f}} \left(1 - \frac{\pi^2}{6k^2} \left(N_1^2 -4N_1N_2+ N_2^2-1\right) \right) +\mathcal{O}\left( k^{-3}\right).
\end{equation}
This result is valid in the non-planar limit, for any value of $N_1$ and $N_2$.
\section{Multiple windings}
\label{sec:multiplewindings}

The expectation value of BPS Wilson loops winding multiple times the circular contour can be computed from localization \cite{Marino:2009jd}, producing exact results valid at framing one. The perturbative analysis performed above at generic framing can be generalized to accommodate multiple windings, providing a more general result depending on two parameters, the winding number $m$ and the framing number $\mathfrak{f}$. The result at $\mathfrak{f}=1$ matches localization, thus confirming the cancellation of the cohomological anomaly.

Multiply wound Wilson loops in perturbation theory were addressed in \cite{Bianchi:2016gpg}. Given a path-ordered integral with $m$ windings emerging from a certain diagram, the approach consists in decomposing the integration contour in such a way to cook up a recursion relation. This requires introducing auxiliary integrals, whose integration domains are not necessarily path-ordered. Also these integrals are subject to recursion relations, thereby providing a complete system of recursive equations. This system is then solved, yielding the multiple winding result in terms of the initial conditions for the various integrals of the system. These are singly wound integrals, not necessarily path-ordered.

In order to make contact to usual path ordered integrals, their integration domains can be decomposed into a union of completely ordered sets at the price of some permutations of the Wilson loop parameters. Such permutations can be undone by relabeling the integration variables so as to attain the usual ordering $2\pi>\tau_1>\tau_2>\dots>0$. In such a process, the symmetries of the specific integrands can be used to streamline the identification of the final integrals as combinations of those appearing in perturbation theory at single winding. A simple example of such an algorithm is provided below, whereas more complex applications for two-loop diagrams are showcased in appendix \ref{app:multiplewinding}. Finally, the process lands on relations between diagrams at multiple and single windings.

While this approach was employed in \cite{Bianchi:2016gpg} to successfully test localization results for bosonic Wilson loops up to three loops, the analysis of fermionic operators did not correctly take into account the antiperiodicity of fermionic integrands. We amend those results in this section for the one-loop and two-loop fermionic diagrams.

Starting from the superconnection \eqref{eq:fermioniWL1}, the multiple winding operator is defined as
\begin{equation}
    W^m_{1/2} = \text{STr}\left( \cP\left( e^{-i\int_0^{2\pi m} d\tau \cL}\right) \begin{pmatrix}
        1 & 0 \\ 0 & (-1)^m
    \end{pmatrix} \right),
\end{equation}
where $m$ is an integer counting the number of windings. We avoid normalizing the Wilson loop at multiple windings by its tree level expectation value $\left(N_1 - (-1)^mN_2\right)$ since no simplifications come about.

\paragraph{First order.} At framing 0, the one-loop result for these operators vanishes for any $m$. At framing 1 the result is known from localization and, as for the single winding case, the cohomological equivalence implies that the fermionic diagram must vanish. For odd windings this is trivially verified, since whatever fermionic contribution is originated from the upper-left block of the superconnection will be precisely cancelled by its complex conjugate from the lower-right 
\begin{equation}
    \left.\vcenter{\hbox{\includegraphics[width=0.1\textwidth]{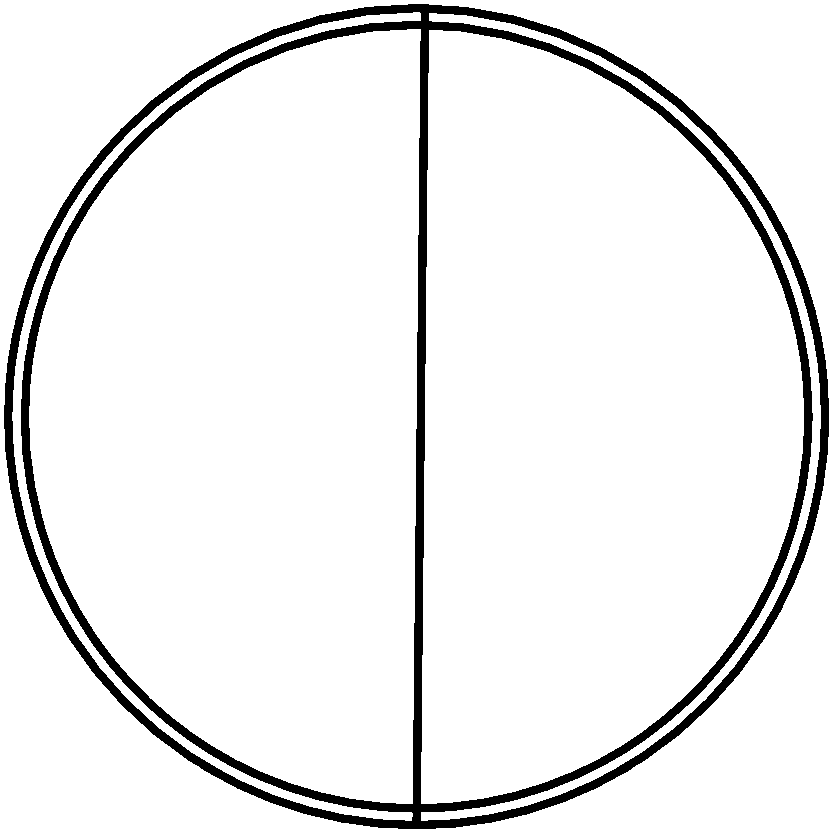}}} \,\,\right|_{m \text{ odd}} = \left.-\frac{i\pi}{k} \,  N_1N_2 \,\mathfrak{f}\,( 1 + (-1)^m )\,\,\right|_{m \text{ odd}} = 0.
\end{equation}
We denote multiple winding contours graphically with a double lined circle. For even windings the contributions from the two blocks of the superconnection sum up. Hence, an explicit evaluation of these terms is required.

At multiple windings, the fermionic exchange at one loop gives rise to an integral of the schematic form
\begin{equation}
I_1(m) = \int_0^{2m\pi} d\tau_1 \int_0^{\tau_1} d\tau_2\, f(\tau_1,\tau_2),
\end{equation}
for some integrand $f(\tau_1,\tau_2)$. Due to the fermionic couplings \eqref{eq:roadmapcouplingswithphases}, such a function is antiperiodic in both its variables: $f(\tau_1 + n_1 2\pi,\tau_2 + n_2 2\pi) = (-1)^{n_1+n_2}\, f(\tau_1,\tau_2)$.
Consequently, the following recursion relations hold
\begin{equation}\label{eq:rsyst1Lf}
    \begin{cases}
        I_1(m) = I_1(m-1) - (-1)^m\, I_{12}(m-1) + I_1(1)\\
        I_{12}(m) = I_{12}(m-1) - (-1)^m I_{12}(1)
    \end{cases},
\end{equation}
where the auxiliary integral $I_{12}(m)$ is defined as
\begin{equation}
    I_{12}(m) \equiv \int_0^{2\pi}d\tau_1\int_0^{2\pi m} d\tau_2\,\,f(\tau_1,\tau_2)\,.
\end{equation}
The derivation of such relations is along the lines of \cite{Bianchi:2016gpg} and rests ultimately on splitting contours, with additional sign factors, emerging when shifting integration variables by $2\pi(m-1)$ to send integrals to the first circle. The system of recursion relations is easily solved by
\begin{equation}
        I_1(m) = \frac{1}{4} \left[(1- (-1)^m - 2m ) I_{12}
        (1) + 4 m\, I_1(1)\right]\,.
\end{equation}
The final solution depends on the initial conditions $I_1(1)$ and $I_{12}(1)$. For the case of an antisymmetric integrand $f(\tau_1,\tau_2)=-f(\tau_2,\tau_1)$, $I_{12}(1)=0$, which implies
\begin{equation}
        I_1(m) = m\, I_1(1),
\end{equation}
with a scaling linear in $m$. Such is the situation for the integral at zero framing. Since in this case the singly wound result vanishes in dimensional regularization, no contribution is generated at any windings.

At generic framing, the part of the fermion couplings sensitive to framing is symmetric under the exchange $x_1\longleftrightarrow x_2$. Such a symmetry is spoiled when splitting contours under the framing procedure, since $x_1$ and $x_2$ explicitly belong to distinct curves. However, it is expected to be restored in the small displacement limit and we will assume that this is the case. This implies that $I_{12}(1)=2\, I_1(1)$, yielding
\begin{equation}
    I_1(m) = \frac{1-(-1)^m}{2}\, I_1(1).
\end{equation}
This result states that the framing-dependent contribution is expected to be the same as for single winding at all odd windings and to vanish at even windings. We corroborated such a finding with extensive numeric checks. This result entails that at even windings the contributions of each separate block from the superconnection vanishes individually.

Summarizing, the fermionic contribution at one loop vanishes for all windings and framing number 
\begin{equation}
    \vcenter{\hbox{\includegraphics[width=0.1\textwidth]{img/1Lfermionm.png}}} = 0.
\end{equation}
At odd windings this is due to opposite contributions from the two blocks. At even windings the two blocks vanish separately. The final result is in full agreement with the localization prediction, requiring fermionic contributions to vanish for all winding numbers.

\paragraph{Second order.} At two loops, the same method applies to the relevant diagrams with more technical complexity. The details of the calculation are presented in appendix \ref{app:multiplewinding} and we only state the final results here. The punchline is that for multiple winding all the fermionic contributions develop the same $m^2$ scaling, combined with a color factor $(N_1-(-1)^m\, N_2)$ depending on the winding.

At framing one, this automatically implies that the same cancellation mechanisms occurring for single winding also occur at multiple winding $m$. In particular, this enforces the cancellation of fermionic diagrams required for the cohomological equivalence to hold. This means that the localization result is verified explicitly for multiple windings as well. 

From the results of appendix \ref{app:multiplewinding}, we can supply a two-loop expression for the unnormalized expectation value of the 1/2 BPS fermionic Wilson loops at generic framing $\mathfrak{f}$ and winding $m$ (including non-planar terms)
\begin{align}\label{eq:genericfm}
\vev{W_{1/2}^m}_{\mathfrak{f}} =& 
N_1 -(-1)^m N_2 +
\frac{i\pi}{k}\, m^2 (N_1^2 + (-1)^m N_2^2) \mathfrak{f} \cr &+ 
\frac{\pi ^2}{6k^2}\, m^2 \Big[
N_1^2 (3 N_2-N_1)-(-1)^m N_2^2 (3 N_1-N_2)
\cr &
-\mathfrak{f}^2 \left(N_1-(-1)^mN_2\right)\left(\left(2 m^2+1\right) \left( N_1^2+(-1)^m N_1 N_2 + N_2^2\right) -3 N_1 N_2
\right)\cr &
+\left(1- \mathfrak{f}^2 \left(m^2-1\right)\right) \left(N_1 -(-1)^m N_2\right)
\Big] +\mathcal{O}\left(k^{-3}\right),
\end{align}
In the relevant limits, $m\to1$, $\mathfrak{f}\to1$ or $\mathfrak{f}\to0$, the expression reproduces correctly the known results from localization and the perturbative analysis of \cite{Bianchi:2013zda,Bianchi:2013rma,Griguolo:2013sma}. Framing does not exponentiate, as is the case for multiple windings in Chern-Simons theory \cite{Brini:2011wi}.

\section{Conclusions}
\label{sec:6}

We have performed the perturbative evaluation -- up to two loops -- of the 1/2 BPS fermionic Wilson loop of ABJ(M) theory at generic framing. This is necessary to directly match the matrix model prediction \eqref{eq:MMresult}, which is valid for $\mathfrak{f}=1$, from perturbation theory. Previous computations at framing zero \cite{Bianchi:2013zda,Bianchi:2013rma,Griguolo:2013sma} required the {\it a posteriori} introduction of a framing phase to compensate for a mismatch. Our computation can then be seen as the first direct test of localization for this operator. Specifically, we have computed the framing-dependent contributions coming from fermionic diagrams, which are crucial to reproduce the localization result. The contributions coming from the gauge field, which are expected to arise from the similarity with the pure Chern-Simons case, are in fact not enough, by themselves, to achieve the agreement. It is then interesting to note that in ABJ(M) framing percolates from the Chern-Simons terms to the matter sector of the theory.

The evaluation of the 1/6 BPS bosonic operator at generic framing in \cite{Bianchi:2016yzj} revealed that, at three loops, the framing exponent acquires the non-trivial correction \eqref{eq:1/6BPSframingrelation}, implying that in Chern-Simons-matter theories the framing factor is indeed a non-trivial function with a smooth expansion in the coupling constants $N_1/k, N_2/k$. It would be interesting to exploit the techniques developed in this paper to attempt a three-loop evaluation of the 1/2 BPS Wilson loop. A non-trivial contribution to framing is expected, which would correct the identity in \eqref{eq:MMrelations} at this order.   

A natural generalization of this project is the perturbative evaluation at generic framing of the interpolating Wilson loops introduced in \cite{Castiglioni:2022yes}. Those operators preserve just one supercharge of the theory and are therefore 1/24 BPS. They are obtained as a certain ${\cal Q}$-deformation of the 1/6 BPS bosonic Wilson loop \eqref{eq:bosonicWL1} and are thus guaranteed to be in the same cohomological class (and to be computed by the same matrix model insertion). Not surprisingly however, the perturbative evaluation performed in \cite{Castiglioni:2022yes} at framing zero yields a result that depends on the interpolating parameters (the $\alpha(\bar\alpha)$ and $\beta(\bar\beta)$ in the formula below), even though these do not enter the insertion \eqref{eq:MM12},
\begin{equation}
\vev{W_{1/24}}_{\mathfrak{f}=0}=
1-\frac{\pi^2}{6 k^2}\left[
N_1^2-4 N_1 N_2+N_2^2 -3 N_1 N_2 (\bar\alpha \alpha-\beta \bar\beta-1)^2
\right]
+{\cal O}\left(k^{-3}\right).
\end{equation}

A phase can be introduced \cite{Castiglioni:2022yes} to account for the different framing in the two prescriptions and obtain agreement with the matrix model prediction. On the other hand, performing the computation at framing $\mathfrak{f}=1$ from the beginning, as done here for the 1/2 BPS fermionic Wilson loop, is expected to eliminate the parameter dependence and give \eqref{eq:MMresult} without the need for compensating phases. This is in fact what we observe in a preliminary analysis, which is complicated by the need to take into account the renormalization of the parameters under RG flow. We plan to report on these results soon~\cite{WIP}.

Similarly, one could study the 1/6 BPS fermionic latitude \eqref{eq:latitudecontour}-\eqref{eq:latitudecouplings} at generic framing and verify the expected cohomological equivalence at $\mathfrak{f}=\nu$. In section \ref{sec:firstorder} we have done this at first order in perturbation theory, obtaining (here we include the contribution of the second node and normalize)
\begin{equation}
    \langle W^{(\nu)}_{1/6} \rangle_\mathfrak{f} = 1 + \frac{2\pi i}{k} \frac{N_1 N_2}{N_1 e^{-\frac{i\pi\nu}{2}}-N_2 e^{\frac{i\pi\nu}{2}}} (\nu - \mathfrak{f}) \cos\frac{\pi\nu}{2}+{\cal O}\left(k^{- 2}\right)\,.
\end{equation}
The result is in agreement with the expectation that cohomological equivalence should hold at $\mathfrak{f}=\nu$ and a vanishing first-order correction \cite{Bianchi:2018bke}. It would be interesting to push this result to higher order in perturbation theory.

Finally, another challenging direction to pursue would be to understand what is the physical meaning of framing in the defect CFT living on the Wilson loop, and what is, if any, its holographic dual. 


\section*{Acknowledgements}
MB is supported by Fondo Nacional de Desarrollo Cient\'ifico y Tecnol\'ogico, through Fondecyt Regular 1220240 and Fondecyt Exploraci\'on 13220060. LC, SP, MT and DT are supported in part by the INFN grant {\it Gauge and String Theory (GAST)}. DT would like to thank FAPESP’s partial support through the grant 2019/21281-4.

\newpage 

\appendix
\section{Gauss linking integral}
\label{app:calugareanu}

In this appendix we prove that the Gauss linking integrand in \eqref{eq:linking}, when multiplied by a generic analytic function, still evaluates to framing $\mathfrak{f}$ in the $\delta \to 0$ limit. This has been used in the computation of the single fermion exchange in section \ref{sec:firstorder}. 

We start with a brief review of the Gauss linking integral. In three-dimensional Euclidean space, let $\Gamma$ be a closed curve defined by coordinates $x^\mu (\tau)$, with $0\leq \tau \leq a$. An infinitesimally deformed curve $\Gamma_\mathfrak{f}$ can be defined as
\begin{equation}
\label{eq:x2contour}
   \Gamma_\mathfrak{f} :  x^{\mu} \to x^{\mu}+n^{\mu}(\tau) \, \delta  \,, \qquad |n(\tau)|=1\,,
\end{equation}
where $\delta$ is a small deformation parameter and $n^\mu(\tau)$ is a vector field orthonormal to $\Gamma$. For non-intersecting curves, the Gauss linking integral
\begin{equation}
\label{eq:linkingApp}
    \chi(\Gamma,\Gamma_\mathfrak{f}) = \frac{1}{4\pi}\int_{\Gamma}dx_1^{\mu} \int_{\Gamma_\mathfrak{f}} dx_2^{\nu} \, \epsilon_{\mu\nu\rho} \frac{(x_1-x_2)^{\rho}}{|x_1-x_2|^3}\,,
\end{equation}
provides a topologically invariant quantity that captures the number of coils of $\Gamma_\mathfrak{f}$ around $\Gamma$. For smooth deformations of $\Gamma$ and $\Gamma_\mathfrak{f}$, \eqref{eq:linkingApp} remains valid as long as the resulting contours remain non-intersecting. This is the case, for instance, for the circle \eqref{eq:maximalcircle}, the latitude \eqref{eq:latitudecontour} and the toroidal helices \eqref{eq:2loopParametrization} considered in the body of the paper.

The integral in \eqref{eq:linkingApp} does not depend on $\delta$. Therefore, one can safely take the limit of coincident curves, thus obtaining what is called {\em framing} $\mathfrak{f}$. Precisely,
\begin{equation}\label{eq:caluga}
    \mathfrak{f}\equiv \lim_{\delta\to 0}\chi(\Gamma,\Gamma_\mathfrak{f}) = \chi(\Gamma) + 2 \int_\Gamma d\tau\, {\rm T}\,,
\end{equation}
where 
$\chi(\Gamma)$ is the self-linking number of $\Gamma$, defined as
\begin{equation}
\label{eq:cotorsion}
    \chi(\Gamma) = \frac{1}{4\pi}\int_{\Gamma}dx_1^{\mu} \int_{\Gamma} dx_2^{\nu} \, \epsilon_{\mu\nu\rho} \frac{(x_1-x_2)^{\rho}}{|x_1-x_2|^3}\,,
\end{equation}
and T is the torsion associated to the curve $\Gamma$. The torsion appears as a consequence of the non-commutativity of the $\delta\to0$ limit with the integrations along the curves. In particular, while $\mathfrak{f}$ is topologically invariant, $\chi(\Gamma)$ itself is not, and requires the torsion to compensate for its metric dependence. An explicit proof of \eqref{eq:caluga} can be found in \cite{calugareanu1959integrale}.

Now we prove that if the integrand appearing in \eqref{eq:linkingApp} is multiplied by a generic analytic function $h(\tau_1-\tau_2)\equiv h(\tau_{12})$, in the $\delta \to 0$ limit it still evaluates to  framing. Precisely, the following identity holds 
(here $x_i \equiv x(\tau_i)$)
\begin{align}\label{eq:toprove}
\begin{split}
          \lim_{\delta\to0} \; \frac{1}{4\pi h_0} \int_\Gamma d\tau_1 \int_{\Gamma_\mathfrak{f}} d\tau_2\,  \dot{x}_1^\mu \dot{x}_2^\nu \, \epsilon_{\mu\nu\rho}\frac{(x_{1}-x_2)^\rho}{|x_1-x_2|^{3}} \, h(\tau_{12}) = \mathfrak{f} \,,
\end{split}
\end{align}
where $h_0$ is the zero mode of $h$ in its series expansion. 

In order to prove \eqref{eq:toprove} we first insert the mode expansion of $h(\tau_{12})$, splitting the zero mode from the rest
\begin{align}\label{eq:toprove2}
\begin{split}  
\lim_{\delta\to0} \, \int_{\Gamma_\mathfrak{f}} d\tau_2 \int_\Gamma d\tau_1 \, & \dot{x}_1^\mu \dot{x}_2^\nu \,  \epsilon_{\mu\nu\rho}\frac{(x_{1}-x_2)^\rho}{|x_1-x_2|^{3}} \, h(\tau_{12}) = 4 \pi h_0\, \mathfrak{f} \\
         & + \lim_{\delta\to0} \, \sum_{m \geq 1} h_m \int_{\Gamma_\mathfrak{f}} d\tau_2\,  \int_\Gamma d\tau_1 \, \dot{x}_1^\mu \dot{x}_2^\nu \,   \epsilon_{\mu\nu\rho}\frac{(x_1-x_2)^\rho }{|x_1-x_2|^{3}} \, (\tau_1-\tau_2)^m \,.
\end{split}
\end{align}
We then prove that the second line is zero, {\it i.e.} the Gauss linking integrand multiplied by any positive  power $(\tau_1-\tau_2)^m$ vanishes in the  $\delta\to0$ limit.

\begin{figure}[ht]
    \centering
    \includegraphics[width=0.35\textwidth]{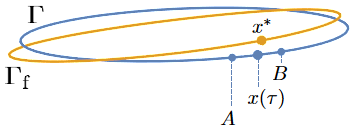}
    \caption{Along the framed curve $\Gamma_\mathfrak{f}$, the point $x^*$ is fixed. Along the circle $\Gamma$, one defines its projection $x(\tau)$ and the neighbouring points $A=x(\tau+ H)$ and $B=x(\tau-H)$.}
    \label{fig:calug}
\end{figure}

To this end, we restrict to the case of circular contours.  We consider $\Gamma$ to be the unit circle and $\Gamma_\mathfrak{f}$ its framed version, as in figure \ref{fig:calug}. 
Going along the lines of \cite{calugareanu1959integrale}, on $\Gamma_\mathfrak{f}$ we fix a point $x^*$ such that $x(\tau_1 = \tau)$ is the projection of $x^*$ onto the circle. We then pick up a small neighbourhood of $\tau$ on $\Gamma$, $AB \equiv (\tau-H, \tau+H)$ with $H$ being a small number, and split the $\tau_1$ integration along $\Gamma$ as
\begin{equation}
\label{eq:integralsplit}
    \int_\Gamma d\tau_1 \longrightarrow \int_{\Gamma-AB} d\tau_{1} +
    \int_{AB} d\tau_{1}\,.
\end{equation}

Integrals appearing in the second line of \eqref{eq:toprove2} can then be performed by fixing $x_2 = x^*$ and focusing on the $\tau_1$ integration first, for which we use the splitting \eqref{eq:integralsplit}. Having removed the $\tau$ neighbourhood, the integral along $\Gamma-AB$ is well defined and finite, and goes to zero for $\delta \to 0$, for any $H \neq 0$ fixed.\footnote{This is true in general for any planar contour.} It is then sufficient to focus on the integral over $AB$, spanned by  $\tau-H < \tau_1 < \tau +H$, which contains the potentially singular point $\tau_1=\tau$. 

For $H$ small, we expand the integrand at low orders in $(\tau_1-\tau) \equiv t \delta$ ($|t\delta| < H$). First, we write 
\begin{align}\label{eq:expandedterms}
    x^* = x(\tau) +  n(\tau) \, \delta \simeq
    x_1 + \left(n_1 -\dot{x}_1\,t \right) \delta   +  f_1(t\delta) \, \delta + f_2(t\delta)\, , 
\end{align}
where $f_1, f_2$ encode the rest of the expansions of $n(\tau)$ and $x(\tau)$ respectively, with $f_1(t\delta) \sim \cO(t\delta)$ and $f_2(t\delta) \sim \cO(t^2\delta^2)$. They are bounded functions in the interval $|t\delta| < H$. Moreover, since on the circle $\ddot{x}^\mu = - x^\mu$, $f_2$ is a linear combination of $x_1^\mu$ and $\dot{x}_1^\mu$ only.  

Exploiting the expansion \eqref{eq:expandedterms}, the denominator in \eqref{eq:toprove2} approximates to 
\begin{equation}
     \frac{1}{|x_{1}-x^*|^3} = \frac{1}{\delta^3(t^2 + 1)^{3/2}} \left( 1 + \frac{1}{t^2 + 1} \mathcal{O}(t\delta) + \frac{1}{\delta(t^2 + 1)} \mathcal{O}(t^2\delta^2) \right)\,.
\end{equation}
Regarding the expansion of the numerator, for any $m>0$ we have
\begin{equation}\label{eq:firsttermT}
\begin{split}
    (\tau_1-\tau)^m\, \epsilon_{\mu\nu\rho} \, \dot{x}_1^\mu \,
    \dot{x}^{*\nu} 
    \,  (x_1-x^*)^\rho = (t\delta)^m \big( & -\delta^2 \,\epsilon_{\mu\nu\rho} \,  \dot x_1^\mu (\dot{n}_1^\nu - \ddot{x}_1^\nu t ) \, n_1^{\rho}  \\
    & \qquad  + \delta^2 \cO(t\delta) + \delta \cO(t^2\delta^2)  \, \big)\, ,
    \end{split}
\end{equation}
where we have neglected terms which vanish due to the planarity of the contour.

Putting all together and changing integration variable from $\tau_1$ to $t$, we obtain
\begin{equation}\label{eq:intcaluga}
\begin{split}
    \int_{-\frac{H}{\delta}}^{\frac{H}{\delta}} d t \, (t \delta)^m & \left(- \epsilon_{\mu\nu\rho} \, \dot x^\mu_1 \, ( \dot{n}_1^\nu - \ddot{x}_1^\nu t) \, n_1^{\rho} + \cO(t\delta) +  \frac{1}{\delta} \cO(t^2\delta^2) \right) \times \\
    & \left( \frac{1}{(t^2+1)^{3/2}}+  \frac{1}{(t^2+1)^{5/2}}\cO(t \delta) + \frac{1}{\delta(t^2+1)^{5/2}}\cO(t^2 \delta^2) \right)\,.
    \end{split}
\end{equation}
Now we take the $\delta \to 0$ limit  keeping $H$ fixed. The range of integration becomes infinite, and since $t = \frac{\tau_1-\tau}{\delta} \sim \delta^{-1}$, the integrals might be singular in the large $t$ region.  

Recalling that $\cO(\dots)$ represents bounded functions in the $|t\delta| < H$ interval, and that the coefficients of the $t$-expansion are also bounded in this interval, in the integrand we can focus only on the $t$ powers.  Applying power counting term by term in \eqref{eq:intcaluga} we are left with integrals of the form
\begin{equation}
    \delta^p \int_{-\frac{H}{\delta}}^{\frac{H}{\delta}} d t \, \frac{ t^{q} }{(t^2+1)^{3/2}} \sim \mathcal{O}(\delta^{p-q+2})\,,
    \qquad \delta^p \int_{-\frac{H}{\delta}}^{\frac{H}{\delta}} d t \, \frac{ t^{q} }{(t^2+1)^{5/2}} \sim \mathcal{O}(\delta^{p-q+4}),
\end{equation}
for $p,q$ integers. A detailed analysis reveals that all the $t$ integrals appearing in \eqref{eq:intcaluga}  go as $\delta$ to some positive power, for any $m>0$ fixed. Therefore \eqref{eq:intcaluga} goes to zero as $\delta\to0$, for any $m>0$. Consequently, the second line of \eqref{eq:toprove2} does not contribute in the limit, and identity \eqref{eq:toprove} holds on the circle.

The proof can be easily generalized to non-circular but closed, planar contours. In this case extra contributions are present, which go as $H \times \cO(1)$ for $\delta \to 0$. They come from integrating bounded functions of $t\delta$ on the $(-H,H)$ interval. However, these contributions are arbitrarily small and can be neglected when we eventually send $H \to 0$ \cite{calugareanu1959integrale}. 

\vskip 10pt
This general result is useful to determine the  framed integral coming from the one-loop fermion exchange in \eqref{eq:cF}. In fact, writing explicitly one of the two contributions
\begin{align}
\begin{split} \label{eq:ffbar}
        \langle \bar{f}_1f_2\rangle &\sim \int_\Gamma d\tau_1 \int_{\Gamma_\mathfrak{f}} d\tau_2\, \dot{x_1}^\mu \dot{x_2}^\nu \,  \epsilon_{\mu\nu\rho}\frac{(x_{1}-x_2)^\rho}{|x_1-x_2|^{3}}\, e^{i\frac{\tau_{12}}{2}}\,,
\end{split}
\end{align}
we note that the integrand is exactly of the form \eqref{eq:toprove2} with 
$h(\tau_{12})=e^{i\frac{\tau_{12}}{2}}$.  Since in this case $h_0=1$, this contribution is proportional to $4\pi \mathfrak{f}$, with the linking number $\mathfrak{f} \in\mathbb{Z}$ defined in \eqref{eq:caluga}.


\section{Conventions and Feynman rules}
\label{app:abjm}

We work in three-dimensional Euclidean space with coordinates $x^{\mu}=(x^1,x^2,x^3)$. The three-dimensional gamma matrices are defined as
\begin{equation}\label{eq:gamma}
    (\gamma^{\mu})^{ \ \beta}_{ \alpha}=(\sigma^1,\sigma^2,\sigma^3)_\alpha^{\ \beta}\,,
\end{equation}
with $(\sigma^{i} )^{ \ \beta}_{  \alpha}$ ($\alpha,\beta=+,-$) being the Pauli matrices, such that $\gamma^{\mu}\gamma^{\nu}=\delta^{\mu\nu}+i\epsilon^{\mu\nu\rho}\gamma_{\rho}$, where $\epsilon^{123}=\epsilon_{123}=1$ is totally antisymmetric. Spinorial indices are lowered and raised as $(\gamma^{\mu})^{\alpha}_{\ \beta}=\epsilon^{\alpha\gamma}(\gamma^{\mu})^{\ \delta}_{\gamma} \epsilon_{\beta\delta}$, with $\epsilon_{+-}=-\epsilon^{+-}=-1$. The Euclidean action of $U(N_1)_k\times U(N_2)_{-k}$ ABJ(M) theory is
\begin{equation}\label{eq:ABJMaction}
\begin{split}
    S_{\textrm{ABJ(M)}}&=\frac{k}{4\pi} \int d^3 x\,  \epsilon^{\mu\nu\rho}\Big\{ -i\text{Tr}\left( A_{\mu}\partial_{\nu}A_{\rho} +\frac{2i}{3}A_{\mu}A_{\nu}A_{\rho} \right)+i\text{Tr}\left( \hat A_{\mu}\partial_{\nu}\hat A_{\rho} +\frac{2i}{3}\hat A_{\mu}\hat A_{\nu}\hat A_{\rho} \right) \\ & + \text{Tr}\left[ \frac{1}{\xi}(\partial_{\mu}A^{\mu})^2 - \frac{1}{\xi}(\partial_{\mu}\hat A^{\mu})^2 +\partial_{\mu} \bar c D^{\mu}c-\partial_{\mu} \bar{\hat c}D^{\mu}\hat c\right] \Big\} \\ & + \int d^3x \text{Tr}\left[ D_{\mu} C_I D^{\mu} \bar C^I +i\bar\psi^I \gamma^{\mu}D_{\mu} \psi_I \right]\\ & \begin{split}- \frac{2\pi i}{k}\int d^3 x \text{Tr}\Big[& \bar C^I C_I \psi_J \bar\psi^J - C_I \bar C^I \bar\psi^J \psi_J + 2C_I\bar C^J \bar\psi^I\psi_J \\ &-2\bar C^I C_J \psi_I \bar\psi^J - \epsilon_{IJKL}\bar C^I \bar \psi^J \bar C^K \bar \psi^L +\epsilon^{IJKL} C_I \psi_J C_K \psi_L \Big]+ S^{\text{bos}}_{\text{int}}\,,\end{split}
\end{split}
\end{equation}
with covariant derivatives defined as
\begin{equation}
\begin{split}
    &D_{\mu} C_I = \partial_{\mu} C_I +i A_{\mu} C_I -i C_I \hat A_{\mu}\,, \qquad D_{\mu} \bar C^I=\partial_{\mu} \bar C^I -i \bar C^I A_{\mu} + i\hat A_{\mu} \bar C^I\,, \\ & D_{\mu}\bar\psi^I=\partial_{\mu} \bar\psi^I + iA_{\mu} \bar\psi^I - i\bar\psi^I \hat A_{\mu}\,, \qquad D_{\mu} \psi_I =\partial_{\mu}\psi_I -i\psi_I A_{\mu} +i\hat A_{\mu} \psi_I\,.
\end{split}
\end{equation}
We work in Landau gauge for vector fields and in dimensional regularization with $d=3-2\epsilon$. The tree-level propagators are 
\begin{equation}
\label{eqn:propagator}
    \begin{split}
        \langle (A_{\mu})_p^{\ q}(x)(A_{\nu})_r^{\ s}(y)\rangle^{(0)} &=\delta_p^s\delta_r^q\, i \left( \frac{2\pi}{k} \right) \, \frac{\Gamma(\frac{3}{2}-\epsilon)}{2\pi^{\frac{3}{2}-\epsilon}}\frac{\epsilon_{\mu\nu\rho}(x-y)^{\rho}}{|x-y|^{3-2\epsilon}},\\
        \langle (\hat A_{\mu})_{\hat p}^{\ \hat q}(x)(\hat A_{\nu})_{\hat r}^{\ \hat s}(y)\rangle^{(0)} &=-\delta_{\hat p}^{\hat s}\delta_{\hat r}^{\hat q} \, i \left( \frac{2\pi}{k} \right) \, \frac{\Gamma(\frac{3}{2}-\epsilon)}{2\pi^{\frac{3}{2}-\epsilon}}\frac{\epsilon_{\mu\nu\rho}(x-y)^{\rho}}{|x-y|^{3-2\epsilon}},\\
        \langle  (\psi_I^{\alpha})_{\hat i}^j(x) (\bar\psi_{\beta}^J)_k^{\hat l} (y) \rangle^{(0)} & = -i\delta_I^J\delta_i^{\hat l}\delta_k^{j } \frac{\Gamma(\frac{3}{2}-\epsilon)}{2\pi^{\frac{3}{2}-\epsilon}}\frac{(\gamma_{\mu})^{\alpha}_{\ \beta}(x-y)^{\mu}}{|x-y|^{3-2\epsilon}}\\ & =i\delta_I^J\delta_i^{\hat l}\delta_k^{j } (\gamma_{\mu})^{\alpha}_{\ \beta}\partial_{\mu}\left( \frac{\Gamma(\frac{1}{2}-\epsilon)}{4\pi^{\frac{3}{2}-\epsilon}} \frac{1}{|x-y|^{1-2\epsilon}}\right), \\
        \langle (C_I)_i^{\hat j}(x)(\bar C^J)_{\hat k}^l(y)\rangle^{(0)} &= \delta_I^J \delta_i^l \delta_{\hat k}^{\hat j}  \frac{\Gamma(\frac{1}{2}-\epsilon)}{4\pi^{\frac{3}{2}-\epsilon}}\frac{1}{|x-y|^{1-2\epsilon}} ,
    \end{split}
\end{equation}
while the one-loop propagators are
\begin{equation}
\label{eqn:onelooppropagator}
\begin{split}
    \langle (A_{\mu})_i^{\ j}(x) (A_{\nu})_k^{\ l}(y) \rangle^{(1)} &= \delta_i^l\delta_k^j \left(\frac{2\pi}{k}\right)^2 N_1 \frac{\Gamma^2(\frac{1}{2}-\epsilon)}{4\pi^{3-2\epsilon}}\left[ \frac{\delta_{\mu\nu}}{|x-y|^{2-4\epsilon}}-\partial_{\mu}\partial_{\nu}\frac{|x-y|^{2\epsilon}}{4\epsilon(1+2\epsilon)} \right] ,\\
    \langle (\hat A_{\mu})_{\hat i}^{\ \hat j}(x) (\hat A_{\nu})_{\hat k}^{ \ \hat l}(y) \rangle^{(1)} &= \delta_{\hat i}^{\hat l}\delta_{\hat k}^{\hat j} \left(\frac{2\pi}{k}\right)^2 N_2 \frac{\Gamma^2(\frac{1}{2}-\epsilon)}{4\pi^{3-2\epsilon}}\left[ \frac{\delta_{\mu\nu}}{|x-y|^{2-4\epsilon}}-\partial_{\mu}\partial_{\nu}\frac{|x-y|^{2\epsilon}}{4\epsilon(1+2\epsilon)} \right] ,\\
    \langle  (\psi_I^{\alpha})_{\hat i}^j(x) (\bar\psi^J_{\beta})_k^{\hat l}(y) \rangle^{(1)} &= i \delta_I^J \,\delta_{\hat i}^{\hat l}\delta_{\hat k}^{\hat j} \, \delta^{\alpha}_{\beta}\left( \frac{2\pi}{k} \right) (N_1-N_2) \frac{\Gamma^2(\frac{1}{2}-\epsilon)}{16\pi^{3-2\epsilon}}\frac{1}{|x-y|^{2-4\epsilon}}.
\end{split}
\end{equation}
The Latin indices are color indices. For instance, $(A_{\mu})_i^{\ j} \equiv A_\mu^a (T^a)_i^{\ j}$ where $T^a$ are $U(N_1)$ generators in the fundamental representation.


\section{Multiple winding at two loops}
\label{app:multiplewinding}

In this appendix we provide details on two-loop diagrams at multiple windings. As recalled in section \ref{sec:multiplewindings}, multiple winding contributions from purely bosonic diagrams were correctly determined at two loops in \cite{Bianchi_2016}. We report the results here for completeness
\begin{align}
\vcenter{\hbox{\includegraphics[width=0.08\textwidth]{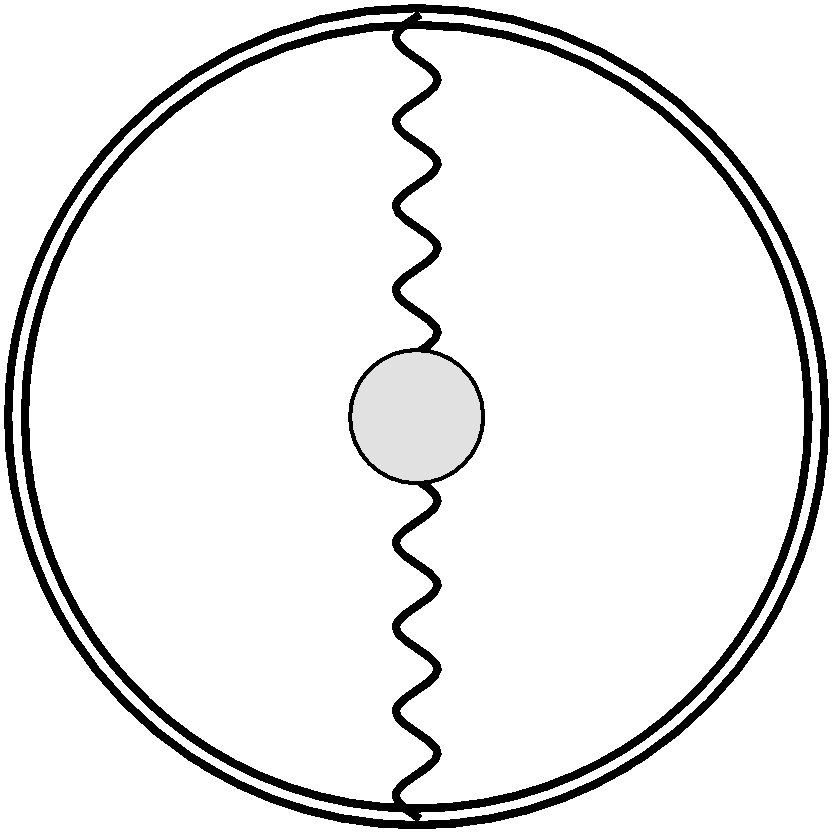}}}+\vcenter{\hbox{\includegraphics[width=0.08\textwidth]{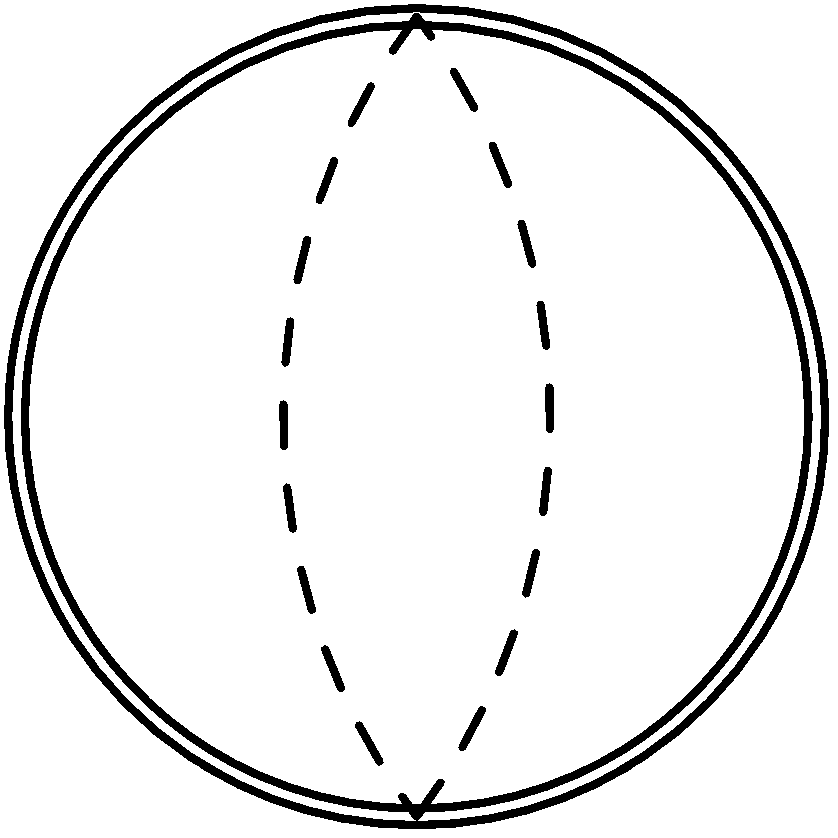}}}=& \; \frac{\pi ^2}{k^2} \, m^2 N_1 N_2 \left(N_1-(-1)^m N_2\right),\nonumber\\
\vcenter{\hbox{\includegraphics[width=0.08\textwidth]{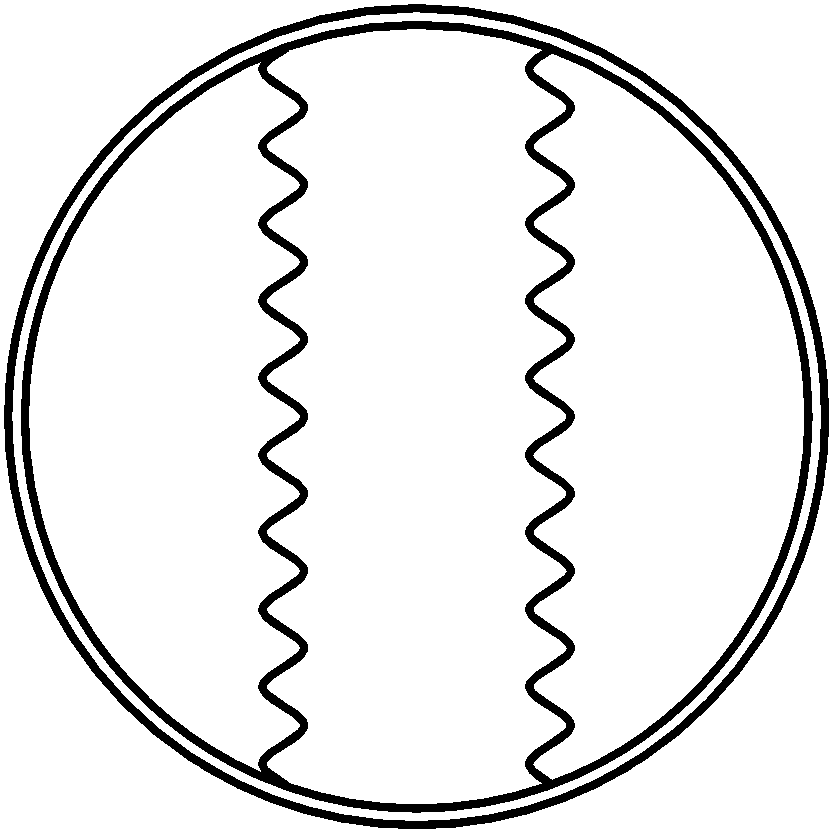}}}=&- \frac{\pi^2}{6k^2} \, m^2 \left(2 m^2+1\right) \left(N_1^3-(-1)^m N_2^3\right) \mathfrak{f}^2 \nonumber\\
& - \frac{\pi^2}{6k^2} \,  m^2 \left(m^2-1\right) \left(N_1 -(-1)^m N_2\right) \mathfrak{f}^2, \nonumber\\
\vcenter{\hbox{\includegraphics[width=0.08\textwidth]{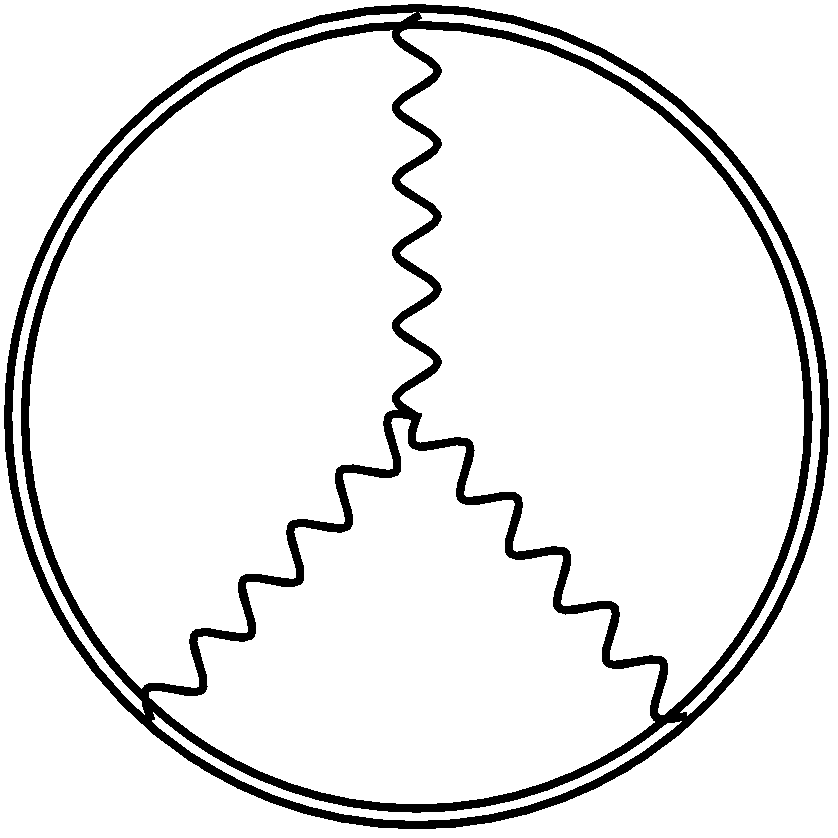}}}=&- \frac{\pi ^2}{6k^2} \, m^2 \left(N_1^3-(-1)^m N_2^3-N_1+(-1)^m N_2\right) . 
\end{align}
The expectation value of the Wilson loop is not normalized for multiple windings.
As a result of the cohomological equivalence, such diagrams evaluated at framing $\mathfrak{f}=1$ reproduce the localization outcome, implying that the sum of the fermionic contributions should vanish at $\mathfrak{f}=1$ for any winding.

Due to the spinor couplings \eqref{eq:roadmapcouplingswithphases}, because of the overall factors $e^{\pm i \frac{\tau}{2}}$, the fermionic integrands are antiperiodic for $2\pi$ shifts of the parameters associated to fermions, irrespective of whether the contribution is framing-dependent or independent. This was not properly taken into consideration in \cite{Bianchi_2016} and here we provide a correct analysis, finally leading to the two-loop expectation value of the 1/2 BPS Wilson loop at generic framing and winding \eqref{eq:genericfm}. The various diagrams are analyzed separately.


\subsubsection*{Double fermion exchange}

The double fermion exchange diagram exhibits an antiperiodic integrand in each variable 
\begin{equation}
    F_1(m) \equiv \int_0^{2\pi m} d\tau_1 \int_0^{\tau_1} d\tau_2\int_0^{\tau_2} d\tau_3\int_0^{\tau_3} d\tau_4 \,\,f(\tau_1,\tau_2,\tau_3,\tau_4),
\end{equation}
where
\begin{equation}
    f(\tau_1+2\pi n_1,\tau_2+2\pi n_2,\tau_3+2\pi n_3,\tau_4+2\pi n_4)=(-1)^{n_1+n_2+n_3+n_4}f(\tau_1,\tau_2,\tau_3,\tau_4).
\end{equation}
A system of recursion relations can be constructed
\begin{equation}
  \begin{cases}
    F_1(m) = F_1(m-1) - (-1)^m F_{1,4}(m-1)+F_{1,3}(m-1)-(-1)^m F_{1,2}(m-1)+F_1(1) 
\\
F_{1,2}(m) = F_{1,2}(m-1)+F_{1,2,4}(m-1)-(-1)^m F_{1,2,3}(m-1)-(-1)^m F_{1,2}(1)
\\
F_{1,3}(m)=F_{1,3}(m-1)-(-1)^m F_{1,3,4}(m-1)+F_{1,3}(1)
\\
F_{1,4}(m)=F_{1,4}(m-1)-(-1)^m F_{1,4}(1)
\\
F_{1,2,3}(m)=F_{1,2,3}(m-1)-(-1)^m F_{1,2,3,4}(m-1)+F_{1,2,3}(1)
\\
F_{1,2,4}(m)=F_{1,2,4}(m-1)-(-1)^m F_{1,2,4}(1)
\\
F_{1,3,4}(m)=F_{1,3,4}(m-1)-(-1)^m F_{1,3,4}(1)
\\
F_{1,2,3,4}(m)=F_{1,2,3,4}(m-1)-(-1)^m F_{1,2,3,4}(1),
\end{cases}
\end{equation}
with the definitions
\begin{equation}\label{eq:mwintdef}
    F_{i_1, i_2, \dots, i_k}(m) \equiv \int_{\mathrm{D}}\,\,
    \prod_{l=1}^n dt_l
    \,\, f(\{\tau_j\}).
\end{equation}
The integration domain is 
\begin{align}
    \mathrm{D}=&
    2\pi>\tau_{i_{1}}>\tau_{i_1+1}>\dots>\tau_{i_2-1}>0 \,\cup \,2\pi>\tau_{i_{2}}>\tau_{i_2+1}>\dots>\tau_{i_3-1}>0 \,\cup \,\dots \nonumber\\&\cup\, 2\pi m>\tau_{i_{k}}>\tau_{i_{k}+1}>\dots>t_{n}>0,
\end{align}
namely, the indices $i_1, \dots, i_{k-1}$ indicate integrations up to $2\pi$, the last index $i_k$ indicates an integration up to $2\pi m$ and the remaining integrations are path-ordered. The solution of the recursion relations for $F_1(m)$ reads
\begin{align}
    F_1(m) =&
    m\, F_1(1)+
    \frac{1}{4} \left(-2 m-(-1)^m+1\right)\, F_{1,2}(1)+
    \frac{1}{2} (m-1) m\, F_{1,3}(1)+\nonumber\\&+
    \frac{1}{4} \left(-2 m-(-1)^m+1\right)\, F_{1,4}(1)+
    \frac{1}{8} \left(-2 (m-2) m+(-1)^m-1\right)\, F_{1,2,3}(1)\nonumber\\&
    -\frac{1}{4} \left((-1)^m-1\right) \left(m+(-1)^m\right)\, F_{1,2,4}(1)+
    \frac{1}{8} \left(-2 (m-2) m+(-1)^m-1\right)\, F_{1,3,4}(1)\nonumber\\&+
    \frac{1}{16} \left(2 m \left(m+(-1)^m-3\right)-3 (-1)^m+3\right)\, F_{1,2,3,4}(1).
\end{align}
This expression is completely general, however it hampers an explicit comparison with single winding contributions, because the domains of integration are not generically path-ordered. In order to reconstruct the original single winding integrals, we can decompose the domains into completely ordered sets and use the symmetries of the integrand (if any) to rename parameters and identify the original contributions.

Let us start with the double fermion exchange from the upper-left block only. In that case, $f(\tau_1,\tau_2,\tau_3,\tau_4)=  - \langle \bar f_1 f_4 \rangle \langle \bar f_3 f_2 \rangle + \langle\bar f_2 f_1 \rangle \langle \bar f_4 f_3 \rangle$. Each fermionic contraction can be split according to \eqref{eq:effectivefermionpropagator2} into a part which is framing-independent and a part which is framing-dependent, which would vanish for planar contours. The framing-independent part $G_{ij}$ is real and antisymmetric under the exchange of the endpoints. The part $F_{ij}$ which would vanish on the plane possesses a symmetric, imaginary part which contributes to framing at one loop and a real, antisymmetric part which does not contribute to framing at one loop and vanishes in the limit of small contour displacement. The real/imaginary and (anti)symmetry properties just emerge from the overall $e^{i\frac{\tau_1-\tau_2}{2}}$ factors from the fermion couplings, multiplying an imaginary and symmetric function. According to such a separation we obtain
\begin{equation}
    f(\tau_1,\tau_2,\tau_3,\tau_4)= \left( \frac{2\pi}{k}\right)^2
    \left(
    - (G_{14}+F_{14})(G_{32}+F_{32}) + (G_{21}+F_{21})(G_{43}+F_{43})
    \right).
\end{equation}
We only retain the part of $F_{ij}$ giving rise to framing, in which case the function becomes symmetric and imaginary. Expanding the contributions, we also use the antisymmetry of $G_{kl}$ and we discard, as in the main calculation, the mixed $F_{ij}\times G_{kl}$ combinations, which would give rise to imaginary two-loop contributions. We are left with
\begin{align}
    f(\tau_1,\tau_2,\tau_3,\tau_4)&= \left( \frac{2\pi}{k}\right)^2
    \left(
    - (G_{14}+F_{14})(-G_{23}+F_{23}) + (-G_{12}+F_{12})(-G_{34}+F_{34})
    \right) \nonumber\\&= \left( \frac{2\pi}{k}\right)^2 \left(G_{14}G_{23} +G_{12}G_{34} -F_{14}F_{23}+F_{12}F_{34}\right).
\end{align}
The various contributions possess different relative signs, hampering symmetrization and factorization. Hence, we analyze them separately. We define:
\begin{align}
    &G^{(ij|kl)}(m) \equiv \int_0^{2\pi m}d\tau_1\int_0^{\tau_3}d\tau_2\int_0^{\tau_2}d\tau_3\int_0^{\tau_3}d\tau_4\, G_{ij}G_{kl},\nonumber\\&
    F^{(ij|kl)}(m) \equiv \int_0^{2\pi m}d\tau_1\int_0^{\tau_3}d\tau_2\int_0^{\tau_2}d\tau_3\int_0^{\tau_3}d\tau_4\, F_{ij}F_{kl}.
\end{align}
The recursion relations are the same, but the initial conditions change for the various contributions. They are a bit lengthy and we will not write them here extensively. The solution of the recursion relations for the various pieces reads
\begin{align}
    G^{(12|34)}(m)  
    =& \frac{1}{2}\Big( \left(2 m^2+(-1)^m-1\right) G^{(13|24)}(1)\cr &
    +2 m^2 \left(G^{(12|34)}(1)+G^{(14|23)}(1)\right)+G^{(14|23)}(1)\left((-1)^m-1\right)\Big),\cr
 G^{(14|23)}(m)
    =& 
    \frac{1}{2} \left((1-(-1)^m) \left(
    G^{(13|24)}(1)+G^{(14|23)}(1)
    \right)-2 m^2 G^{(13|24)}(1)\right),  
    \cr
    F^{(12|34)}(m) 
    =& \frac{1}{2}  \left(1-(-1)^m\right) F^{(12|34)}(1),
    \cr
    F^{(14|23)}(m) =&
    m^2 F^{(14|23)}(1) +\frac{1}{2} \left(1-(-1)^m-2 m^2\right)F^{(12|34)}(1).
\end{align}
For the purely antisymmetric part, corresponding to zero framing, we obtain
\begin{equation}
     G^{(12|34)}(m)+G^{(14|23)}(m)  
    = m^2\left( G^{(12|34)}(1)+G^{(14|23)}(1) \right).
\end{equation}
For the purely symmetric part, which sources framing, we obtain
\begin{equation}
     F^{(12|34)}(m)-F^{(14|23)}(m) 
    = m^2\left( F^{(12|34)}(m)-F^{(14|23)}(m) \right),
\end{equation}
meaning that the framing contribution at multiple winding is $m^2$ times the contribution at single winding. The relative sign difference is crucial to achieve this result. The conclusion is that this block acquires an overall $m^2$ factor altogether. Combining with the other block, the total factor becomes $N_1 N_2 m^2 (N_1-(-1)^m N_2)$ times the single winding contribution
\begin{equation}
    \vcenter{\hbox{\includegraphics[width=0.1\textwidth]{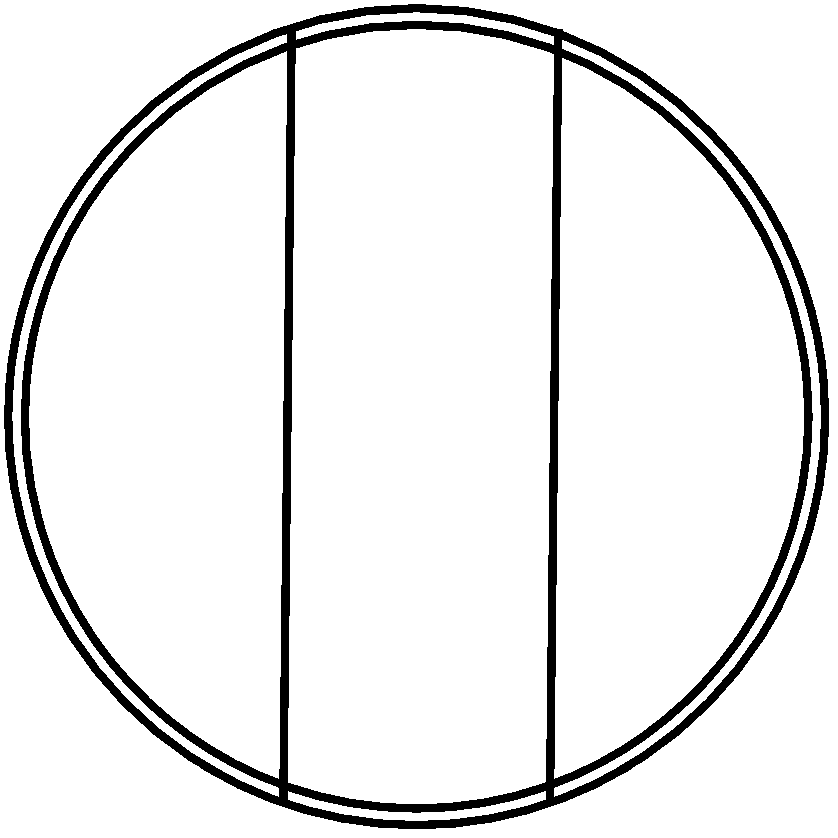}}} =   \frac{\pi^2}{2k^2} \, N_1N_2 (N_1-(-1)^m N_2)(3 - \mathfrak{f}^2) \, m^2,
\end{equation}
where the cartoon stands for the sum of all double fermion exchange diagrams.


\subsubsection*{Mixed fermion-gauge double exchange}

We start from expression \eqref{eq:mixedgauge2} for the the $N_1^2N_2$ contribution. In the case of multiple windings we obtain
\begin{equation}
    N_1^ 2N_2 \left( \langle A_1 A_2 \rangle F_{34} + \langle 
    A_3 A_4 \rangle F_{12} +  \langle A_1 A_4 \rangle F_{23} + (-1)^m  \langle A_2 A_3 \rangle F_{41} \right)\, ,
\end{equation}
where the last piece comes from the lower-right block of the superconnection and its additional minus sign is due to a fermion commutation. As before, we have split the fermion contribution into an antisymmetric and a symmetric part and only retained the latter, as the former would give rise to an imaginary piece in the two-loop result. The symmetric part $F_{ij}$ gives rise to the fermionic framing contribution.

We then construct the relevant recursion relations. Though their general structure is the same for all contributions, the relative signs of their terms depend on the specific diagram, since the fermion and gluon exchanges have different periodicity. We report here only their general solutions. We define 
\begin{equation}
    M^{(ij|kl)}(m)\equiv \int_0^{2\pi m}d\tau_1\int_0^{\tau_3}d\tau_2\int_0^{\tau_2}d\tau_3\int_0^{\tau_3}d\tau_4\, F_{ij}\, \langle A_k A_l \rangle.
\end{equation}
The solution to their recursion relations reads
\begin{align}
    M^{(12|34)}(m) =&
    m\, M^{(12|34)}(1)+
    \frac{1}{4} \left(1-2 m-(-1)^m\right)\, \left(M^{(12|34)}_{1,2}(1)+M^{(12|34)}_{1,4}(1)\right)\nonumber\\&+
    \frac{1}{8} \left((-1)^m-2 (m-2) m-1\right)\, \left(M^{(12|34)}_{1,2,3}(1)+M^{(12|34)}_{1,3,4}(1)\right)\nonumber\\&
    -\frac{1}{4} \left((-1)^m-1\right) \left(m+(-1)^m\right)\, M^{(12|34)}_{1,2,4}(1)+
    \frac{1}{2} (m-1) m\, M^{(12|34)}_{1,3}(1) \nonumber\\&+
    \frac{1}{16} \left(2 m \left(m+(-1)^m-3\right)-3 (-1)^m+3\right)\, M^{(12|34)}_{1,2,3,4}(1), \nonumber\\
    M^{(34|12)}(m) =&
    m\, M^{(34|12)}(1)+
    \frac{1}{2} (m-1) m\, \left(M^{(34|12)}_{1,2}(1)+M^{(34|12)}_{1,3}(1)\right) \nonumber\\&+
    \frac{1}{4} \left(-2 m-(-1)^m+1\right)\, M^{(34|12)}_{1,4}(1)+
    \frac{1}{6} (m-2) (m-1) m\, M^{(34|12)}_{1,2,3}(1)\nonumber\\&+
    \frac{1}{8} \left((-1)^m-2 (m-2) m-1\right)\, \left(M^{(34|12)}_{1,2,4}(1)+M^{(34|12)}_{1,3,4}(1)\right)\nonumber\\&+
    \frac{1}{48} \left(-2 (m-2) m (2 m-5)-3 (-1)^m+3\right)\, M^{(34|12)}_{1,2,3,4}(1), \nonumber\\
    M^{(14|23)}(m) =&
    m\, M^{(12|34)}(1)-\frac{1}{4} \left((-1)^m-1\right) \left(m+(-1)^m\right)\, M^{(14|23)}_{1,2,4}(1)\nonumber\\&-
    \frac{1}{4} \left(2 m+(-1)^m\right)\, \left(M^{(14|23)}_{1,4}(1)+M^{(14|23)}_{1,2}(1)\right)+
    \frac{1}{2} (m-1) m\, M^{(14|23)}_{1,3}(1)
    \nonumber\\&
    +
    \frac{1}{8} \left((-1)^m-2 (m-2) m-1\right)\, \left(M^{(14|23)}_{1,3,4}(1)+M^{(14|23)}_{1,2,3}(1)\right)\nonumber\\&+
    \frac{1}{16} \left(2 m \left(m+(-1)^m-3\right)-3 (-1)^m+3\right)\, M^{(14|23)}_{1,2,3,4}(1), \nonumber\\
    M^{(23|14)}(m) =&
    m\, M^{(23|14)}(1)+
    \frac{1}{2} (m-1) m\, \left(M^{(23|14)}_{1,2}(1)+M^{(23|14)}_{1,4}(1)\right)\nonumber\\&+
    \frac{1}{8} \left(-2 (m-2) m+(-1)^m-1\right)\, \left(M^{(23|14)}_{1,2,3}(1)+M^{(23|14)}_{1,3,4}(1)\right)\nonumber\\&+
    \frac{1}{4} \left(1-2 m-(-1)^m\right)\, M^{(23|14)}_{1,3}(1)+
    \frac{1}{6} (m-2) (m-1) m\, M^{(23|14)}_{1,2,4}(1) \nonumber\\&+\frac{1}{48} \left(-2 (m-2) m (2 m-5)-3 (-1)^m+3\right)\, M^{(23|14)}_{1,2,3,4}(1).
\end{align}
The notation for the integrals follows \eqref{eq:mwintdef}.
Again, these expressions are completely general, but they obscure the reference to the original single winding integrals, which can be recovered by splitting integration contours into completely ordered domains and using symmetries:
\begin{align}
    M^{(12|34)}(m)   
    =& 
    \frac{1}{2} m (m-1)\, M^{(34|12)}(1)+
    \frac{1}{2} (m+1) m\, M^{(12|34)}(1)\nonumber\\&+
    \frac{1}{4} \left(2 m^2+(-1)^m-1\right)\, \left(M^{(23|14)}(1)+
    \frac12\, M^{(13|24)}(1)+\frac12\,M^{(24|13)}(1)\right),
    \nonumber\\
    M^{(34|12)}(m)  
    =& 
    \frac{1}{2} m (m+1)\, M^{(34|12)}(1)+
    \frac{1}{2} (m-1) m\, M^{(12|34)}(1)\nonumber\\&+
    \frac{1}{4} \left(2 m^2+(-1)^m-1\right)\, \left(M^{(23|14)}(1)+
    \frac12\, M^{(13|24)}(1)+\frac12\,M^{(24|13)}(1)\right),\nonumber\\
M^{(14|23)}(m)   
    =&
    \frac{1}{4} (-1)^m \left(-2 m^2+(-1)^m-1\right)M^{(14|23)}(1)-\frac{1}{4} (-1)^m\left(2 m^2+(-1)^m-1\right)\nonumber\\&
     ~~~~\times \left( M^{(12|34)}(1)+M^{(34|12)}(1)+M^{(23|14)}(1)+M^{(13|24)}(1)+M^{(24|13)}(1)\right),\nonumber\\
M^{(23|14)}(m)   
    =& 
    \frac{1}{4} \left(2 m^2+(-1)^m-1\right) \left( M^{(12|34)}(1)+M^{(34|12)}(1)\right)\nonumber\\&+
    \frac{1}{4} \left(2 m^2-(-1)^m+1\right)\, M^{(23|14)}(1)
    +\frac{1}{4} \left(-2 m^2-(-1)^m+1\right)M^{(14|23)}(1).
\end{align}
Considering the combination which enters the $N_1^2N_2$ contribution, we obtain for multiple windings 
\begin{align}
    N_1^2N_2 \left( M^{(12|34)}(m) +M^{(34|12)}(m) + M^{(23|14)}(m) + (-1)^m M^{(14|23)}(m)\right) \nonumber\\
    =N_1^2N_2 m^2\left( M^{(12|34)}(1) +M^{(34|12)}(1) + M^{(23|14)}(1) - M^{(14|23)}(1) \right),
\end{align}
meaning that the multiple winding contribution is precisely $m^2$ times the contribution at winding 1. The $N_1N_2^2$ term differs by an overall $-(-1)^m$ factor. The sum over the two blocks gives a factor $N_1N_2(N_1-(-1)^m N_2)m^2$ times the upper-left $m=1$ contribution
\begin{equation}\label{eq:mixedResultmw}
\vcenter{\hbox{\includegraphics[width=0.08\textwidth]{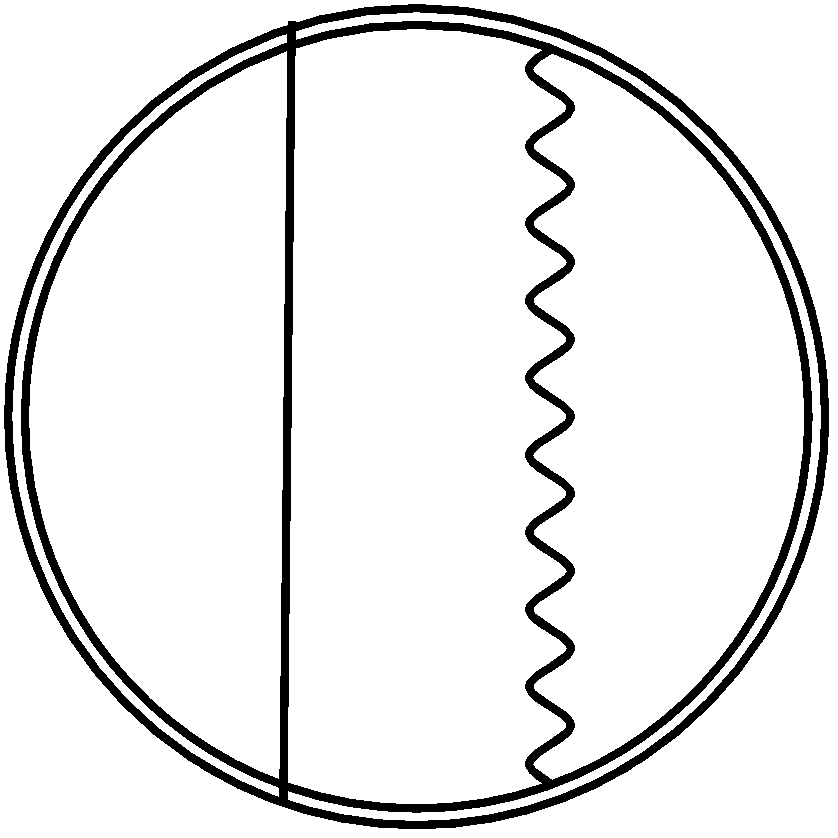}}} =  \frac{\pi^2}{k^2} N_1N_2 (N_1-(-1)^m N_2) \, m^2 \,
 \mathfrak{f}^2  \,.
\end{equation}


\subsubsection*{One-loop corrected fermion exchange}

The integral for such a diagram is framing-independent, therefore we evaluate it on a circular contour, where the contribution from each block is proportional to
\begin{equation}
I_2(m) = \int_0^{2\pi m}d\tau_1\int_0^{\tau_1} d\tau_2\,\,\frac{\cos\frac{\tau_1-\tau_2}{2}}{\left(\sin^2\frac{\tau_1-\tau_2}{2}\right)^\alpha} ,  
\end{equation}
with $\alpha=1-2\epsilon$.
This is similar to the single fermion contribution, except that the integrand is now symmetric. As derived in section \ref{sec:multiplewindings} for the fermion framing contribution at one loop, the solution of the recursion reads
\begin{equation}
    I_2(m) = \frac{1-(-1)^m}{2}\, I_2(1),
\end{equation}
so it vanishes at even loops. At odd loops it is equal to the single winding case, where it is vanishing after summing over the upper-left and lower-right blocks. In conclusion, this diagram vanishes identically for all windings, though via different mechanisms at odd and even windings.


\subsubsection*{Vertex diagram}

Each vertex contribution can be represented as an integrand of the form $\langle f_1\bar f_2 A_3\rangle \rightarrow v(\tau_1,\tau_2|\tau_3)$. The first two entries represent the parameters of the fermion insertions and the last the gluon. Therefore $v$ is antiperiodic in its first two arguments and periodic in the last one
\begin{equation}
v(\tau_1+2\pi n_1,\tau_2+2\pi n_2|\tau_3+2\pi n_3) = (-1)^{n_1+n_2}v(\tau_1,\tau_2|\tau_3).
\end{equation}
Performing the expansion of the Wilson loop we obtain symbolically for the $N_1^2N_2$ color contribution
\begin{equation}\label{eq:vertexv}
    N_1^2 N_2 \left( v(\tau_1,\tau_2|\tau_3) + v(\tau_2,\tau_3|\tau_1) + (-1)^m\, v(\tau_3,\tau_1|\tau_2) ,\right)
\end{equation}
the last piece coming from the lower-right block and the extra minus sign due to a fermion commutation. The corresponding path-ordered integrals at generic winding read
\begin{equation}
    V^{(ij|k)}_1(m)\equiv \int_0^{2\pi m} d\tau_1 \int_0^{\tau_1} d\tau_2\int_0^{\tau_2} d\tau_3\, v(\tau_i,\tau_j|\tau_k).
\end{equation}
Recursion relations can be set for such integrals. For instance, for $V^{(12|3)}_1(m)$ we obtain
\begin{equation}
    \begin{cases}
    V^{(12|3)}_1(m)=V^{(12|3)}_1(m-1)+V^{(12|3)}_{1,3}(m-1)-(-1)^m V^{(12|3)}_{1,2}(m-1)+V^{(12|3)}_1(1)\\
    V^{(12|3)}_{1,2}(m)=V^{(12|3)}_{1,2}(m-1)-(-1)^m
   (V^{(12|3)}_{1,2,3}(m-1)+V^{(12|3)}_{1,2}(1))\\
    V^{(12|3)}_{1,3}(m)=V^{(12|3)}_{1,3}(m-1)+V^{(12|3)}_{1,3}(1)\\
   V^{(12|3)}_{1,2,3}(m)=V^{(12|3)}_{1,2,3}(m-1)+V^{(12|3)}_{1,2,3}(1),
   \end{cases}
\end{equation}
with analogous index definition as in \eqref{eq:mwintdef}. Then, the solution of the recursion relations for the relevant contributions reads
\begin{align}
    V^{(12|3)}_1(m) =& 
    m\, V^{(12|3)}_1(1)
    +\frac12 (m-1) m\, V^{(12|3)}_{1,3}(1)
    -\frac14 \left(2 m+(-1)^m-1\right)\, V^{(12|3)}_{1,2}(1)\nonumber\\&
    + \frac18 \left(-2 (m-2)m+(-1)^m-1\right)\, V^{(12|3)}_{1,2,3}(1),
\nonumber\\
    V^{(23|1)}_1(m) =& 
    m\, V^{(23|1)}_1(1)
    +\frac12 (m-1) m\, V^{(23|1)}_{1,2}(1)
    -\frac14 \left(2 m+(-1)^m-1\right)\,  V^{(23|1)}_{1,3}(1)\nonumber\\&
    +\frac18 \left(-2 (m-2) m+(-1)^m-1\right)\,  V^{(23|1)}_{1,2,3}(1),
\nonumber\\
    V^{(13|2)}_1(m) =& 
    m\, V^{(13|2)}_1(1)
    -\frac{1}{4}(-1)^m \left(V^{(13|2)}_{1,3}(1)+V^{(13|2)}_{1,2}(1)+(m-1)\, V^{(13|2)}_{1,2,3}(1)\right)\nonumber\\&
    -\frac14 (2m-1)\, \left(V^{(13|2)}_{1,3}(1)+V^{(13|2)}_{1,2}(1)\right)
    +\frac{1}{4} (m-1)\, V^{(13|2)}_{1,2,3}(1).
\end{align}
The initial conditions depend crucially on whether the integrand is symmetric or antisymmetric in the exchange of the first two arguments. As in the single winding calculation of section \ref{sec:circWLperturbative}, we only focus on the framing-independent part. Following \cite{Bianchi:2013rma,Griguolo:2013sma}, the integrand at framing 0 is manifestly antisymmetric in the exchange of the fermion endpoints. After decomposing into path-ordered contours and upon using this antisymmetry we find
\begin{align}
    V^{(12|3)}_1(m) =& \frac{1}{2} m (m+1) V^{(12|3)}_1(1)+\frac{1}{2}m (m-1)
   V^{(23|1)}_1(1)
    \nonumber\\&
   +\frac{1}{4}\left(2 m^2+(-1)^m-1\right) V^{(13|2)}_1(1),\nonumber\\
    V^{(23|1)}_1(m) =& \frac{1}{2}m (m-1) V^{(12|3)}_1(1)+\frac{1}{2} m (m+1)
   V^{(23|1)}_1(1)\nonumber\\&+\frac{1}{4}\left(2 m^2+(-1)^m-1\right) V^{(13|2)}_1(1),\nonumber\\
    V^{(13|2)}_1(m) =& -\frac{1}{2} \left((-1)^m-1\right) V^{(13|2)}_1(1).
\end{align}
When summed over the three pieces \eqref{eq:vertexv}, taking only the antisymmetric part of $v$ in its first two entries, it gives
\begin{align}
    &N_1^2 N_2 \left( V^{(12|3)}_1(m) + V^{(23|1)}_1(m) - (-1)^m\, V^{(13|2)}_1(m) \right)  \nonumber\\&~~~~~~~~~ = N_1^2 N_2 m^2 \left( V^{(12|3)}_1(1) + V^{(23|1)}_1(1) + V^{(13|2)}_1(1) \right),
\end{align}
therefore the multiple winding scaling for the framing 0 term is $m^2$. Summing contributions proportional to the color factor $N_1N_2^2$ we obtain the full result
\begin{equation}
    \vcenter{\hbox{\includegraphics[width=0.1\textwidth]{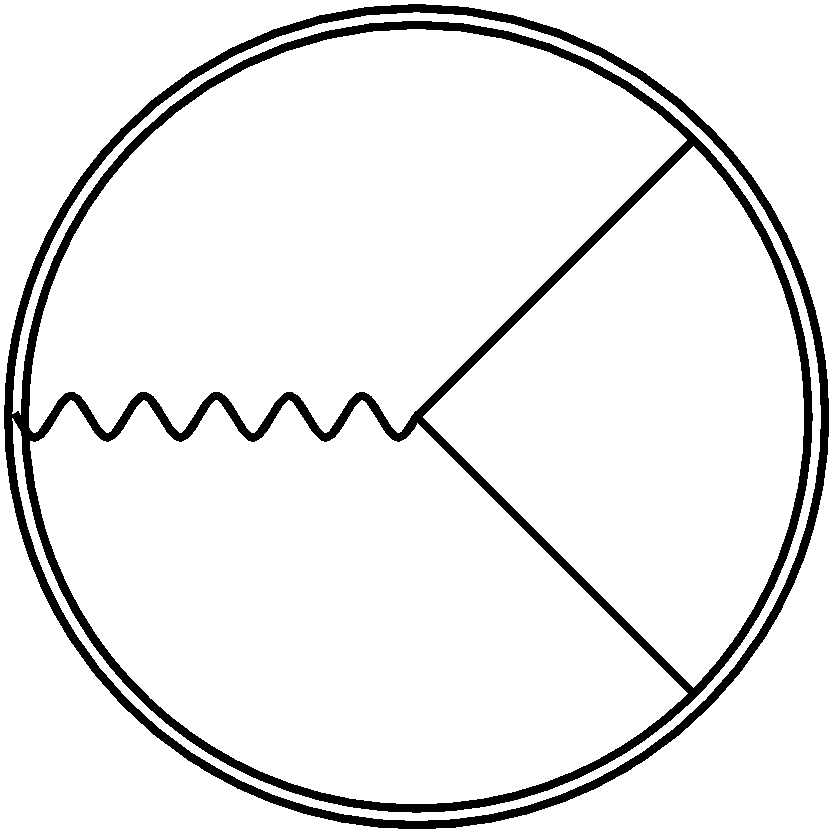}}}= -\frac{2\pi^2}{k^2}\, N_1N_2 (N_1-(-1)^m N_2) \, m^2 .
\end{equation}

\newpage

\bibliographystyle{JHEP}
\bibliography{refs}
\newpage 

\end{document}